\documentclass[12pt]{iopart}

\usepackage[mathscr]{eucal} %  use \EuScript (\mathcal unchanged)
\usepackage[dvips]{graphicx}

\usepackage{etoolbox}
\makeatletter
\newcommand{\mainmatter}{%
  \setcounter{footnote}{0}%
  \patchcmd{\@makefntext}{\fnsymbol}{\arabic}{}{}%
  \patchcmd{\@thefnmark}{\fnsymbol}{\arabic}{}{}%
  \def\@makefnmark{\textsuperscript{\arabic{footnote}}}%
}
\makeatother

\expandafter\let\csname equation*\endcsname\relax
\expandafter\let\csname endequation*\endcsname\relax
\usepackage{amsmath,amsfonts,amssymb,amsthm,amscd}

\usepackage{longtable}
\usepackage{array}
\usepackage{float}
\usepackage[utf8]{inputenc}
\usepackage[colorlinks=true, urlcolor=blue,linkcolor=blue,citecolor=blue]{hyperref}

\numberwithin{equation}{section}

\newtheorem{theorem}{Theorem}[section]

\theoremstyle{definition}

\newtheorem{example}[theorem]{Example}
\newtheorem{remark}[theorem]{Remark}

\newcommand{\Z}{{\mathbb Z}}

\newcommand{\C}{{\mathbb C}}

\begin{document}

\title[]{Generalized hydrodynamics in box-ball system}

\author{Atsuo Kuniba$^1$, Gr\'egoire Misguich$^{2,3}$ and Vincent Pasquier$^2$}

\address{$^1$ Institute of Physics, 
University of Tokyo, Komaba, Tokyo 153-8902, Japan}
\address{$^2$ Institut de Physique Th\'eorique, 
Universit\'e Paris Saclay, CEA, CNRS, F-91191 
Gif-sur-Yvette, France}
\address{$^3$ Laboratoire de Physique Th\'eorique et Mod\'elisation, CNRS UMR 8089,
  CY Cergy Paris Universit\'e 95302 Cergy-Pontoise Cedex, France.}

\eads{\mailto{atsuo.s.kuniba@gmail.com}, \mailto{gregoire.misguich@ipht.fr}
\mailto{vincent.pasquier@ipht.fr}}

%\vspace{0.5cm}
%\begin{center}{\bf Abstract}\end{center}
\begin{abstract}
Box-ball system (BBS) is a prominent example of
integrable cellular automata in one dimension connected to 
quantum groups, Bethe ansatz, ultradiscretization, tropical geometry and so forth.
In this paper we study the generalized Gibbs ensemble of BBS soliton gas
by thermodynamic Bethe ansatz and generalized hydrodynamics. 
The results include the solution to the speed equation for solitons, 
an intriguing connection of the effective speed with the period matrix of
the tropical Riemann theta function, an explicit description of the 
density plateaux that emerge from domain wall initial conditions
including their diffusive corrections.
\end{abstract}
%\submitto{\jpa}

% From the IOP guidemines:
% Keywords are required for all submissions. Authors should supply a minimum of three
% (maximum seven) keywords appropriate to their article
\noindent{\it Keywords\/}: Box-ball system, Integrable cellular automata, 
Solitons gas, Generalized Gibbs ensemble, Generalized hydrodynamics

%\maketitle
\mainmatter
%--------------------------------------------------
\section{Introduction}
%--------------------------------------------------

The box-ball system (BBS) invented originally in \cite{TS90} 
is an integrable cellular automaton on one dimensional lattice.
It accommodates solitons exhibiting factorized scattering.
Here is an example of collision of three solitons with amplitude 4, 2 and 1,
where time evolution goes downward:

\vspace{0.1cm}
\begin{center}
$ \ldots  00111100000110000010000000000000000000000000000
\ldots $
$ \ldots 00000011110001100001000000000000000000000000000
\ldots $
$ \ldots 00000000001110011100100000000000000000000000000
\ldots $
$ \ldots 00000000000001100011011100000000000000000000000
\ldots $
$ \ldots 00000000000000011000100011110000000000000000000
\ldots $
$ \ldots 00000000000000000110010000001111000000000000000
\ldots $
$ \ldots 00000000000000000001101000000000111100000000000
\ldots $
$ \ldots 00000000000000000000010110000000000011110000000
\ldots $
$ \ldots 00000000000000000000001001100000000000001111000
\ldots $
\end{center}

\vspace{0.1cm}
One observes that the larger solitons are faster and a two body 
collision entails a phase shift in the trajectory.
They come back precisely in the original amplitude 1,2 and 4 
after nontrivial intermediate states.

Subsequent studies have clarified that BBS originates  
either in the ultradiscretization of a discrete KdV equation or also in the $q=0$ limit of 
solvable vertex models associated with the quantum group $U_q(\widehat{sl}_2)$.
As such, it inherits and synthesizes a number of rich aspects both in 
classical and quantum integrable systems.
For example, one can construct commuting family of time evolutions 
and the associated conserved quantities
by introducing the {\it carriers} on a hidden auxiliary space 
(see Example \ref{ex:3s}).
Solitons in BBS yield the quasi-particles 
that are exactly counted by Bethe's formula \cite{Be}  and its variant
(see (\ref{ff})) for the number of string solutions to the Heisenberg chain
(soliton/string correspondence).
The equation of motion takes the Hirota-Miwa bilinear form in which 
the role of tau functions is played by 
an ultradiscrete analogue of corner transfer matrices \cite{Bax}.
The initial value problem on a periodic lattice 
is solved by tropical Riemann theta function\footnote{Analogue of the Riemann theta
function in tropical geometry where infinite sum 
is replaced by a minimum or maximum over 
an infinite set.} whose period matrix and Poincar\'e cycle are simply related 
to the Bethe ansatz data, and so forth. 
For the integrability and various generalizations of BBS,  
see for example the review \cite{IKT12} and the references therein.

Recently there has been a renewed interest on BBS from the perspectives of 
statistical physics and probability theory in and out of equilibrium
\cite{CKST18, CS19, CS20, FG18, FNRW, LLP17, LLPS19, KL20}.
Our aim in this paper is to explore such features further in the light of 
generalized hydrodynamics (GHD).
The approach has flourished widely for the Bethe ansatz integrable systems in general 
\cite{Castro-Alvaredo2016, Bertini2016,D19} by developing and unifying the  
ingredients known earlier in 
\cite{Zakharov_1971, Tsarev_1991, Kamchatnov_nonlinear_2000, 
El_2003} for example.

Let us digest the main results of the paper.
We consider the generalized Gibbs ensemble (GGE) of BBS soliton gas and 
apply the  thermodynamic Bethe ansatz (TBA) \cite{YY69,Ta99}.
Due to the soliton/string correspondence mentioned above, 
the TBA equation, the stationary condition of GGE,  
becomes the well-known Y-system involving many temperatures 
(called driving terms) but 
without a spectral parameter.
We present its solution in the form of multi-fugacity series expansion
by invoking the generalized Q-system \cite{KNT02}. 

One of the basic ingredients in GHD is a speed equation.
It governs the effective speed of solitons taking the influence of 
the other ones into account.
We find that the speed equation of BBS is nothing but the 
inversion relation of the tropical period matrix mentioned previously.
The solution, i.e. the effective speed, is thereby identified 
with an appropriately scaled off-diagonal elements 
of the inverse of the tropical period matrix.
These facts are integrated into a quite general proof that the 
current by solitons coincides with the time average of the carrier current
over the Poincar\'e cycle.
This intriguing connection deserves a further investigation which we expect to  
yield a deeper insight into GHD.  

We present a general formalism of the GHD \cite{D19} adapted to BBS.
The essential variable is the occupation function, or equivalently the Y-function,
in the TBA playing the role of the normal mode. 
As an application we study the non-equilibrium dynamics starting from 
domain wall initial conditions. 
It is a typical setting in the Riemann problem called {\em partitioning protocol}.
Our numerical analysis demonstrates the formation of plateaux in the density profile 
after the evolution from the i.i.d. random initial states in the non-empty region.
It testifies the ballistic transport of solitons 
where each plateau is filled with those having a selected list of amplitudes. 
The plateaux exhibit slight broadening in their edges due to the diffusive correction
to the ballistic picture. 
A general recipe to calculate the diffusion constant for such a spread has been 
given in \cite{DBD, GHKV}.
We make use of these GHD machinery 
to derive an analytical formula for the positions and heights of the plateaux, 
and moreover the curves that describe their edge broadening.
They agree with the numerical  data excellently.

The layout of the paper is as follows.
In Sec. \ref{sec:bbs}, we recall the basic features of BBS 
in the periodic boundary condition 
including commuting time evolutions $T_1, T_2, \ldots$ 
and a family of conserved quantities.
In Sec. \ref{sec:gge}, we study the GGE of BBS solitons by TBA.
In particular for the two temperature case, 
closed formulas are given for the densities of strings, holes and the energies in
(\ref{epaz})--(\ref{roaz}).
For the general case, the fugacity expansion  solution is presented in \ref{app:A}.
In Sec. \ref{sec:current},
we study the effective speed of solitons 
and the stationary current in the spatially homogeneous setting.
Our speed equation (\ref{spv}) for the time evolution $T_l$ 
coincides with \cite[eq.(11.7)]{FNRW} at $l=\infty$. 
It is established quite generally that the current due to solitons 
coincides with the time average of the carrier current \cite[Prop.4.3]{KT10}.  
The proof elucidates a new link between the effective speed 
and the period matrix $B$ (\ref{bmat})
which appeared originally in the tropical Riemann theta function \cite[eq.(4.14)]{KT10}.
For the two temperature GGE, the effective speed and the current are obtained 
in closed forms (\ref{mirei}) and  (\ref{chizuko}). 
The latter nontrivially agrees with an alternative derivation (\ref{npn}) based on a transfer matrix formalism. 
It also reproduces the earlier result in \cite[Lem.3.15]{CKST18} 
for $T_\infty$.
In Sec. \ref{sec:5}, we formulate GHD in a form adapted to BBS
and apply it to the density plateaux generated form 
the domain wall initial conditions. 
We perform an extensive numerical analysis and confirm 
an agreement with high accuracy.
Sec. \ref{sec:discussion} contains a future outlook concerning 
a more general BBS.
\ref{app:tr} presents a transfer matrix formalism of the GGE partition
function.
\ref{app:A} relates the TBA equation to the 
generalized Q-system and provides the multi-fugacity series expansion
formulas.
\ref{app:note29} provides a proof of (\ref{kawa}) which leads to the 
general formula (\ref{vsig}) of the effective speed in terms of the hole density.
\ref{app:current1}  and \ref{app:current2} derive an  
explicit formula for the current in homogeneous case by two different methods.
\ref{app:degenerate} recalls the linearly degenerate hydrodynamic type systems 
from \cite{Tsarev_1991,Kamchatnov_nonlinear_2000}.

\section{Box-ball system}\label{sec:bbs}

In this section we recall the basic features of 
the generalized periodic box-ball system (BBS)
equipped with commuting time evolutions $T_1, T_2, \ldots$ \cite{KTT06} which  
includes $T_\infty$ studied in \cite{YYT}.  

\subsection{Combinatorial R}
For a positive integer $l$, set 
$B_l=\{(x_0,x_1) \in (\Z_{\ge 0})^2 \mid  x_0+x_1=l \}$. 
We shall use $\otimes$ to denote the product of the sets  
$B_l$'s and their elements instead of $\times$  
having the crystal structure in mind although its consequence will not be 
utilized explicitly in this paper. 
Define a map $R_{l,m}: B_l \otimes B_m \rightarrow B_m \otimes B_l$ by
\begin{align}
R_{l,m}: & \,(x_0,x_1) \otimes (y_0,y_1) 
\mapsto (\tilde{y}_0, \tilde{y}_1) \otimes 
(\tilde{x}_0, \tilde{x}_1),
\label{rlm}
\\
\tilde{x}_i &=x_i +\min(x_{i+1},y_i)-\min(x_{i},y_{i+1}),
\;\;
\tilde{y}_i = y_i - \min(x_{i+1},y_i)+\min(x_{i},y_{i+1}),
\end{align}
where all the indices are in $\Z_2$.
The map $R_{l,m}$ is known as the combinatorial R of $A^{(1)}_1$.
It is a bijection and satisfies the 
inversion relation $R_{l,m}R_{m,l}=\mathrm{id}_{B_m \otimes B_l}$ and the 
Yang-Baxter relation
$(1\otimes R_{k,l})(R_{k,m}\otimes 1) (1\otimes R_{l,m})
= (R_{l,m}\otimes 1)(1\otimes R_{k,m}) (R_{k,l} \otimes 1)$.
It also enjoys the symmetry corresponding to the Dynkin diagram automorphism of 
$A^{(1)}_1$: 
\begin{align}\label{rw}
R(\omega \otimes \omega) = (\omega \otimes \omega) R,\qquad
\omega\bigl((x_0,x_1)\bigr) = (x_1,x_0).
\end{align}

The BBS considered in this paper is mostly concerned with 
$R_{l,1}: B_l \otimes B_1 \rightarrow B_1 \otimes B_l$.
Its action is given by $(l-n, n) \otimes (1-\eta, \eta) 
\mapsto (1-\tilde{\eta}, \tilde{\eta}) \otimes (l-\tilde{n}, \tilde{n})$,
where $\tilde{\eta}, \tilde{n}$ are determined from $n, \eta$ 
according to the following diagrams:
\begin{align}\label{vd}
\begin{picture}(150,110)(-40,-85)
\put(-130,-30){
\put(18,15.3){$\eta$}
\put(20,10){\vector(0,-1){21}}
\put(-4,-3){$n$}\put(6,0){\vector(1,0){28}}\put(39,-3){$\tilde{n}\qquad=$}
\put(18,-21){$\tilde{\eta}$} 
}
\put(18,14){0}
\put(20,10){\vector(0,-1){21}}
\put(-4,-3){$n$}\put(6,0){\vector(1,0){28}}\put(39,-3){$n\!-\!1$}
\put(18,-22.5){1} 
\put(70,-22){$(n>0)$}
\put(160,0){
\put(18,15){0}
\put(20,10){\vector(0,-1){21}}
\put(-3,-3){$0$}\put(6,0){\vector(1,0){28}}\put(38,-3){0}
\put(18,-22.5){0} 
}
\put(0,-60){
\put(18,15){1}
\put(20,10){\vector(0,-1){21}}
\put(-4,-3){$n$}\put(6,0){\vector(1,0){28}}\put(39,-3){$n\!+\!1$}
\put(18,-22.5){0} 
\put(70,-22){$(n<l)$}
}
\put(160,-60){
\put(18,15){1}
\put(20,10){\vector(0,-1){21}}
\put(-2,-3){$l$}\put(6,0){\vector(1,0){28}}\put(38,-3){$l$}
\put(18,-22.5){1} 
}
\end{picture}
\end{align}

\subsection{States, time evolutions and energies}

The BBS is a dynamical system on a periodic
lattice of size $L$. 
An element  $\eta= \eta_1 \otimes \cdots \otimes \eta_L 
\in B_1^{\otimes L}$ is called a state.
In what follows a local state 
$(1-\eta_i, \eta_i) \in B_1$ is flexibly identified with 
$\eta_i \in \{0,1\}$ and interpreted as a box containing $\eta_i$ balls at site $i \in \Z_L$.
Thus the state $\eta$ is also presented as an array 
$\eta= (\eta_1, \ldots, \eta_{L}) \in \{0,1\}^L$.

In order to introduce the time evolution, 
consider the composition of $R_{l,1}$ $L$ times  
which sends $B_l \otimes B_1^{\otimes L}$ to 
$B_1^{\otimes L} \otimes B_l$.
If $\xi_l \otimes \eta_1 \otimes \cdots \otimes \eta_L \mapsto 
\tilde{\eta}_1 \otimes \cdots \otimes \tilde{\eta}_L \otimes \tilde{\xi}_l$ 
under this map, the situation is depicted as a concatenation of the diagrams 
(\ref{vd}) in the form of a row transfer matrix:
 \begin{align}\label{tl}
\begin{picture}(150,45)(0,-9)
\put(-10,11){$\xi_l$} \put(125,10){$\tilde{\xi}_l$}
\put(12,27){$\eta_1$}\put(32,27){$\eta_2$}
\multiput(0,0)(20,0){2}{\put(15,22){\vector(0,-1){20}}}
\put(12,-8){$\tilde{\eta}_1$}\put(32,-8){$\tilde{\eta}_2$}
\put(0,13){\line(1,0){50}}\put(55,10.5){$\cdots$}
\put(70,13){\vector(1,0){50}}
\put(76,27){$\eta_{L-1}$}\put(102,27){$\eta_L$}
\multiput(70,0)(20,0){2}{\put(15,22){\vector(0,-1){20}}}
\put(76,-8){$\tilde{\eta}_{L-1}$}\put(102,-8){$\tilde{\eta}_L$}
\end{picture}
\end{align}
As with $\eta_i$, we identify the elements $(l-n,n) \in B_l$
attached to the horizontal line with $n \in\{0,1,\ldots, l\}$, and 
regard it as the capacity $l$ {\em carrier} containing $n$ balls \cite{TM97}.
In this interpretation, the diagrams (\ref{vd}) for $R_{l,1}$ 
describe a loading/unloading process of balls between a local box and  
the capacity $l$ carrier which is proceeding to the right. 

Given $\xi_l \in B_l$ and  
$\eta= (\eta_1, \ldots, \eta_{L})$,
the diagram (\ref{tl}) determines
$\tilde{\eta} = (\tilde{\eta}_1, \ldots, \tilde{\eta}_{L})$ and 
$\tilde{\xi}_l \in B_l$ uniquely.
This relation will be expressed as\footnote{
We use $\simeq$ to signify that the RHS is actually 
the result of successive application of $R_{l,1}$ to the LHS
bearing in mind that it causes the isomorphism of crystals
$B_l \otimes B_1^{\otimes L} \simeq B_1^{\otimes L} \otimes B_l$.}   
$\xi_l \otimes \eta \simeq \tilde{\eta} \otimes \tilde{\xi}_l$.

Suppose $\sum_{i \in \Z_L}\eta_i <  L/2$, i.e., the 
ball density is less than one half.
(See Remark \ref{re:01} for the other case.)
Then there is a unique $\xi_l \in B_l$ such that $\tilde{\xi}_l = \xi_l$
in (\ref{tl}).
Denoting it by $c_l(\eta) \in B_l$, it can be produced as
$(l,0) \otimes \eta \simeq \tilde{\eta} \otimes c_l(\eta)$.
These facts have been proved in 
\cite[eqs.(2.9)--(2.11)]{KTT06}.
Based on them we define the time evolution 
$T_l(\eta) =(\eta'_1, \ldots, \eta'_{L})$ 
and the associated $l$-th {\em energy}
$E_l(\eta) \in \Z_{\ge 0}$ of $\eta$ by 
\begin{align}
c_l(\eta) \otimes \eta &\simeq T_l(\eta) \otimes c_l(\eta),
\label{tldef}
\\
E_l(\eta) &= \sum_{i \in \Z_L} \theta(\eta_i >\eta'_i).
\label{el}
\end{align}
Here and in what follows we use
\begin{align}\label{teta}
\theta(\text{true})=1,\qquad \theta(\text{false})=0.
\end{align}
The RHS of (\ref{el}) depends on $l$ via $\eta'_i = T_l(\eta)_i$.
Then the commutativity $T_l T_k(\eta) = T_kT_l(\eta)$
and the energy conservation 
$E_l(T_k(\eta)) = E_l(\eta)$ are valid for all $l,k \in \Z_{\ge 1}$
\cite[Th.2.2]{KTT06}.
The time evolution $T_l$ can be identified with a fusion transfer matrix 
with $l+1$-dimensional auxiliary space at $q=0$ \cite{IKT12}.
The above properties are reminiscent of commuting transfer matrices 
in the sense of Baxter \cite{Bax}. 

When $l=1$, the $R_{l,1}$ is the identity map
$(1-n,n) \otimes (1-\eta_i,\eta_i) \mapsto (1-n,n) \otimes (1-\eta_i, \eta_i)$.
From this fact we see that $T_1$ is a cyclic shift and $E_1$ is a sum of 
a simple nearest neighbor correlation:
\begin{align}
T_1\bigl((\eta_1, \eta_2, \ldots, \eta_L)\bigr) 
&= (\eta_L, \eta_1, \ldots, \eta_{L-1}),
\\
E_1\bigl((\eta_1, \eta_2, \ldots, \eta_L)\bigr) 
&= \sum_{i \in \Z_L}\theta(\eta_i < \eta_{i+1}) 
\label{e1}
\end{align}

As $l \rightarrow \infty$, the time evolution $T_l$ converges to 
some dynamics $T_\infty$ describable without a carrier \cite[Ex.~2.7]{KTT06}.
It is nothing but a translation in the extended affine Weyl group of $A^{(1)}_1$
\cite[Prop.~2.5]{KTT06}, which is referred to as   
Pitman's transformation in probability theory literatures, e.g.,  \cite{CKST18}.

\begin{remark}\label{re:01}
Thanks to the symmetry (\ref{rw}), one can cover the case
$\sum_{i \in \Z_L}\eta_i>L/2$ by 
$T_l(\eta) = \omega^{\otimes L}\bigl( T_l(\omega^{\otimes L}(\eta))\bigr)$.
The energy also obeys 
$E_l(\eta) = E_l(\omega^{\otimes L}(\eta))$.
When $\sum_{i \in \Z_L}\eta_i=L/2$, the $c_l(\eta)$ satisfying (\ref{tldef}) is not unique 
but $T_l(\eta)$ remains unique. 
For the general treatment, see \cite{KTT06}.
\end{remark}

\subsection{Solitons}
We assume $\sum_{i \in \Z_L}\eta_i <  L/2$.
Let us illustrate the solitons along examples.
\begin{example}\label{ex:1s}
For the state $\eta$ of length $L=13$ on the top line, 
its time evolution $T_3^t(\eta)$ (left) and  $T_5^t(\eta)$ (right) are displayed.
\begin{equation*}
\begin{picture}(200,61)(0,50)
\multiput(-40,0)(0,0){1}{
\put(0,99)
{$t=0: \quad 0\; 0\; 0\; 1\; 1\; 1\; 1\; 0\; 0\; 0\; 0\; 0\; 0 $}
\put(0,88)
{$ t=1: \quad  0\; 0\; 0\; 0\; 0\; 0\; 1\; 1\; 1\; 1\; 0\; 0\; 0 $}
\put(0,77)
{$ t=2:  \quad 0\; 0\; 0\; 0\; 0\; 0\; 0\; 0\; 0\; 1\; 1\; 1\; 1 $}
\put(0,66)
{$ t=3:  \quad 1\; 1\; 1\; 0\; 0\; 0\; 0\; 0\; 0\; 0\; 0\; 0\; 1 $}
\put(0,55)
{$ t=4:  \quad 0\; 0\; 1\; 1\; 1\; 1\; 0\; 0\; 0\; 0\; 0\; 0\; 0 $}}

\multiput(135,0)(0,0){1}{
\put(0,99)
{$0\; 0\; 0\; 1\; 1\; 1\; 1\; 0\; 0\; 0\; 0\; 0\; 0$}
\put(0,88)
{$0\; 0\; 0\; 0\; 0\; 0\; 0\; 1\; 1\; 1\; 1\; 0\; 0$}
\put(0,77)
{$1\; 1\; 0\; 0\; 0\; 0\; 0\; 0\; 0\; 0\; 0\; 1\; 1 $}
\put(0,66)
{$0\; 0\; 1\; 1\; 1\; 1\; 0\; 0\; 0\; 0\; 0\; 0\; 0$}
\put(0,55)
{$0\; 0\; 0\; 0\; 0\; 0\; 1\; 1\; 1\; 1\; 0\; 0\; 0$}}
\end{picture}
\end{equation*}
These are examples of one soliton states.
The consecutive array of balls keeps the pattern  
and proceeds to the right periodically 
with the (bare) speed $3$ and $4$.
It is an analogue of stable wave packet.
The energy spectrum $(E_1,E_2,E_3, \ldots)$ is
$(1,2,3,4,4,\ldots)$.
\end{example}

As Example \ref{ex:1s} indicates, 
consecutive $k$ balls ($1$'s) behave as 
a soliton of amplitude $k$, or simply a $k$-soliton in general.
Its speed under the time evolution $T_l$ is $\min(l,k)$.
It does not exceed $l$ since the carrier for $T_l$ can load at most $l$ balls.
A sufficiently isolated $k$-soliton contributes to $E_l$ by $\min(l,k)$.
This suggests us to {\em define} the number $m_k$ of $k$-solitons in a state 
from the conserved energy spectrum as
\begin{align}\label{elm}
E_l = \sum_{k\ge 1}\min(l,k)m_k,\qquad
m_k = -E_{k-1} + 2E_k -E_{k+1}\quad (E_0=0). 
\end{align}
These features are parallel with the BBS on the infinite (non-periodic) lattice
\cite{FOY00}.  
In fact, so defined $m_k=m_k(\eta)$ is know to satisfy 
$m_k\ge 0$ nontrivially \cite[Prop.~3.4]{KTT06}.
The conserved quantity $m=(m_k)_{k \ge 1}$, which is a 
linear recombination of the energy spectrum, is called the 
{\em soliton content} of a state.
It is conveniently depicted as the Young diagram $Y(\eta)$
containing $m_k$ rows of length $k$:
\begin{equation}\label{Yeta}
\begin{picture}(150,88)(0,16)

\put(0,64){$Y(\eta)\;=$}

\put(100,100){\line(-1,0){50}}\put(100,100){\line(0,-1){15}}
\put(85,85){\line(1,0){15}} \put(85,85){\line(0,-1){35}}
\put(96,73){\vector(0,1){10}}
\put(92,64.5){$m_k=m_k(\eta)$}\put(96,60){\vector(0,-1){10}}
\put(85,50){\line(-1,0){10}}\put(75,50){\line(0,-1){20}}
\put(75,30){\line(-1,0){15}}\put(60,30){\line(0,-1){15}}
\put(60,15){\line(-1,0){10}}
\put(50,15){\line(0,1){85}}

\put(62,64.5){\vector(-1,0){11}}\put(64.5, 61){$k$}
\put(72.5,64.5){\vector(1,0){11}}

\end{picture}
\end{equation}
By the definition $Y(\eta)$ is the list of the amplitude of solitons 
which correspond to its rows.
The energy $E_l$ is the number of boxes in the left $l$ columns of $Y(\eta)$.
In particular $E_1$, the number of solitons, is the depth of  $Y(\eta)$, and 
$E_\infty$, the number of balls, is the total area of $Y(\eta)$.
Now it is clear that the energy $E_l$ has the saturation property
\begin{align}\label{ei}
E_1 <  E_2 <  \cdots < E_s =E_{s+1} = \cdots =  E_\infty(\eta) 
= M:=\sum_{i \in \Z_L}\eta_i,
\end{align}
where $s$ is the amplitude of the largest soliton, or equivalently, the width of $Y(\eta)$.
By the definition $m_k=0$ holds for $k>s$. 
The upper bound $M$ is the total number of balls.
Similarly it is known that $T_l(\eta) = T_\infty(\eta)$ if and only if $l \ge s$.

\begin{example}\label{ex:2s}
Let us observe the time evolution $T_4^t(\eta)$ (left) and  $T_5^t(\eta)$ (right) 
of a two soliton state $\eta$ on the top line.  
\begin{equation*}
\begin{picture}(200,60)(0,50)
\multiput(-120,0)(0,0){1}{
\put(0,99)
{$t=0: \quad
1\; 1\; 1\; 1\; 1\; 0\; 0\; 0\; 0\; 1\; 1\; 0\; 0\; 0\; 0\; 0\; 0\; 0\; 0\; 0\; 0\; 0$}
\put(0,88)
{$t=1: \quad
0\; 0\; 0\; 0\; 1\; 1\; 1\; 1\; 1\; 0\; 0\; 1\; 1\; 0\; 0\; 0\; 0\; 0\; 0\; 0\; 0\; 0 $}
\put(0,77)
{$t=2: \quad
0\; 0\; 0\; 0\; 0\; 0\; 0\; 0\; 1\; 1\; 1\; 0\; 0\; 1\; 1\; 1\; 1\; 0\; 0\; 0\; 0\; 0$}
\put(0,66)
{$t=3: \quad
0\; 0\; 0\; 0\; 0\; 0\; 0\; 0\; 0\; 0\; 0\; 1\; 1\; 0\; 0\; 0\; 1\; 1\; 1\; 1\; 1\; 0 $}
\put(0,55)
{$t=4: \quad
1\; 1\; 1\; 0\; 0\; 0\; 0\; 0\; 0\; 0\; 0\; 0\; 0\; 1\; 1\; 0\; 0\; 0\; 0\; 0\; 1\; 1 $}}

\multiput(139,0)(0,0){1}{
\put(0,99)
{$1\; 1\; 1\; 1\; 1\; 0\; 0\; 0\; 0\; 1\; 1\; 0\; 0\; 0\; 0\; 0\; 0\; 0\; 0\; 0\; 0\; 0 $}
\put(0,88)
{$0\; 0\; 0\; 0\; 0\; 1\; 1\; 1\; 1\; 0\; 0\; 1\; 1\; 1\; 0\; 0\; 0\; 0\; 0\; 0\; 0\; 0 $}
\put(0,77)
{$0\; 0\; 0\; 0\; 0\; 0\; 0\; 0\; 0\; 1\; 1\; 0\; 0\; 0\; 1\; 1\; 1\; 1\; 1\; 0\; 0\; 0 $}
\put(0,66)
{$1\; 1\; 0\; 0\; 0\; 0\; 0\; 0\; 0\; 0\; 0\; 1\; 1\; 0\; 0\; 0\; 0\; 0\; 0\; 1\; 1\; 1 $}
\put(0,55)
{$0\; 0\; 1\; 1\; 1\; 1\; 1\; 0\; 0\; 0\; 0\; 0\; 0\; 1\; 1\; 0\; 0\; 0\; 0\; 0\; 0\; 0$}}
\end{picture}
\end{equation*}
A 5-soliton and a 2-soliton are colliding repeatedly and periodically.
Their amplitude 5+2 look 3+4 or 4+3 temporarily 
in the course of the collisions, but 
they do come back to the original 5 and 2 when separated sufficiently.
The collisions cause a shift of the trajectory of free solitons. 
We we call it the {\em phase shift}.
In the above examples, the both larger and smaller solitons have been
dragged to each other by $4$ lattice units.
\end{example}

As observed in Example \ref{ex:2s}, the phase shift in a collision 
of a $j$-soliton and a $k$-soliton is 
$2\min(j, k)$ under any time evolution $T_l$ as long as 
their relative speed $\min(j, l)$ and $\min(k, l)$ are different. 
The independence of the value $2\min(j, k)$ on $l$ is a manifestation of the 
commutativity of the time evolutions.

\begin{example}\label{ex:3s}
Let us observe the time evolution 
$T_2^t(\eta) \in B^{\otimes 19}_1$ of a five soliton state $\eta$ on the top line.
\begin{equation*}
\begin{picture}(200,80)(0,34)
\multiput(-70,0)(0,0){1}{
\put(0,99)
{$t=0: \quad 0\; 1\; 1\; 0\; 0\; 1\; 1\; 0\; 0\; 1\; 1\; 1\; 0\; 0\; 1\; 0\; 1\; 0\; 0 $}
\put(0,88)
{$ t=1: \quad  0\; 0\; 0\; 1\; 1\; 0\; 0\; 1\; 1\; 0\; 0\; 1\; 1\; 1\; 0\; 1\; 0\; 1\; 0$}
\put(0,77)
{$ t=2:  \quad 1\; 0\; 0\; 0\; 0\; 1\; 1\; 0\; 0\; 1\; 1\; 0\; 0\; 1\; 1\; 0\; 1\; 0\; 1 $}
\put(0,66)
{$ t=3:  \quad 1\; 1\; 1\; 0\; 0\; 0\; 0\; 1\; 1\; 0\; 0\; 1\; 1\; 0\; 0\; 1\; 0\; 1\; 0$}
\put(0,55)
{$ t=4:  \quad 0\; 0\; 1\; 1\; 1\; 0\; 0\; 0\; 0\; 1\; 1\; 0\; 0\; 1\; 1\; 0\; 1\; 0\; 1$}
\put(0,44)
{$ t=5:  \quad 1\; 1\; 0\; 0\; 1\; 1\; 1\; 0\; 0\; 0\; 0\; 1\; 1\; 0\; 0\; 1\; 0\; 1\; 0$}}

\put(210,49){
\put(-41,26){$Y(\eta) = $}
\put(0,50){\line(1,0){30}}\put(30,50){\line(0,-1){10}}
\put(0,40){\line(1,0){30}}\put(20,50){\line(0,-1){30}}
\put(0,30){\line(1,0){20}}\put(10,50){\line(0,-1){50}}
\put(0,20){\line(1,0){20}}
\put(0,10){\line(1,0){10}}
\put(0,0){\line(1,0){10}}
\put(0,50){\line(0,-1){50}}
}
\end{picture}
\end{equation*}
The energy spectrum $(E_1,E_2,\ldots)$ is 
$(5,8,9,9,\ldots)$, which corresponds to 
$m_1=m_2=2, m_3=1$ and $m_k=0$ for $k\ge 4$.
The evolution from $T_2(p)$ to $T^2_2(p)$ 
has been computed by the diagram (\ref{tl}) as
\begin{equation*}
\unitlength 0.7mm
{\small
\begin{picture}(130,18)(28,-8)
\put(-1.8,0){
\multiput(0.8,0)(10,0){19}{\line(1,0){4}}
\multiput(2.8,-3)(10,0){19}{\line(0,1){6}}}

\put(0,5){0}
\put(10,5){0}
\put(20,5){0}
\put(30,5){1}
\put(40,5){1}
\put(50,5){0}
\put(60,5){0}
\put(70,5){1}
\put(80,5){1}
\put(90,5){0}
\put(100,5){0}
\put(110,5){1}
\put(120,5){1}
\put(130,5){1}
\put(140,5){0}
\put(150,5){1}
\put(160,5){0}
\put(170,5){1}
\put(180,5){0}

\put(-5,-1){1}
\put(5,-1){0}
\put(15,-1){0}
\put(25,-1){0}
\put(35,-1){1}
\put(45,-1){2}
\put(55,-1){1}
\put(65,-1){0}
\put(75,-1){1}
\put(85,-1){2}
\put(95,-1){1}
\put(105,-1){0}
\put(115,-1){1}
\put(125,-1){2}
\put(135,-1){2}
\put(145,-1){1}
\put(155,-1){2}
\put(165,-1){1}
\put(175,-1){2}
\put(185,-1){1}

\put(0,-7.5){1}
\put(10,-7.5){0}
\put(20,-7.5){0}
\put(30,-7.5){0}
\put(40,-7.5){0}
\put(50,-7.5){1}
\put(60,-7.5){1}
\put(70,-7.5){0}
\put(80,-7.5){0}
\put(90,-7.5){1}
\put(100,-7.5){1}
\put(110,-7.5){0}
\put(120,-7.5){0}
\put(130,-7.5){1}
\put(140,-7.5){1}
\put(150,-7.5){0}
\put(160,-7.5){1}
\put(170,-7.5){0}
\put(180,-7.5){1}

\end{picture}
}
\end{equation*}
\end{example}

%--------------------------------------------------
\section{Generalized Gibbs ensemble of BBS solitons}\label{sec:gge}
%--------------------------------------------------

\subsection{Volume of iso-level set}
We assume $\sum_{i \in \Z_L}\eta_i <  L/2$.
We have seen that the energy spectrum $(E_j)_{j\ge 1}$ (\ref{el}),
the soliton content $(m_j)_{j \ge 1}$ (\ref{elm}) and the 
Young diagram $Y$ (\ref{Yeta}) are equivalent ways of presenting the 
conserved quantities of BBS.
The combination
\begin{align}\label{pj}
p_j = L - 2\sum_{k \ge 1}\min(j,k)m_k = L-2E_j \quad (j \ge 0)
\end{align}
is called the {\em vacancy}, and $p_j/L$ the {\em hole} density.
They will play an important role in the sequel.
From (\ref{ei})  one sees 
$L=p_0 > p_1 >  \cdots  > p_s = p_{s+1} = \cdots  = p_\infty = L-2M\ge 1$.
Given the data $m=(m_1,m_2,\ldots, m_s)\in (\Z_{\ge 0})^s$ such that 
$\sum_{1\le j \le s}j m_j <L/2$,
introduce the set of BBS states whose soliton content is $m$:
\begin{align}
\mathcal{P}_s(m) = \{ \eta=(\eta_1,\ldots, \eta_L)  \in \{0,1\}^L \mid
E_l(\eta) = \sum_{j=1}^s\min(l,j)m_j\;\text{for any}\; l\}.
\end{align}
This is an iso-level set, which is invariant under the time evolutions of BBS.
It consists of solitons with amplitude not exceeding $s$.
From \cite[Cor.~4.4, eq.(4.8)]{KTT06}, its cardinality is given by 
a version of ``Fermionic formula":
\begin{align}
|\mathcal{P}_s(m)| &= 
\frac{L}{L-2M}\prod_{j =1}^s\binom{p_j + m_j-1}{m_j},
\label{ff}
\end{align}
where $M = \sum_{1 \le k \le s}km_k <L/2$ as in (\ref{ei}).
The formula (\ref{ff}) is known valid essentially for the whole 
range $0\le M \le L$. 
In general $m_j=0$ for $j>L/2$, therefore any $s$ such that $s\ge [L/2]$ 
is equivalent and allows all the possible amplitude for the solitons.
See \cite[eqs.(3.6), (4.21)]{KTT06} for the precise description. 

\begin{example}
Cosider $L=9$, $m=(m_1,m_2,m_3,\ldots) = (1,0,1,0,0,\ldots)$,
hence $(p_1,p_2,p_3,\ldots) = (5,3,1,1\ldots)$.
For $s\ge 3$, the iso-level set $\mathcal{P}_s(m)$ consists of the 
45 states given by the $\Z_9$ cyclic shifts of 
\begin{align*}
(000010111),\quad (000100111), \quad (001000111), 
\quad (010000111), \quad (000011011),
\end{align*}
where the last one is the ``intermediate" state during the collision
in which amplitude $3+1$ temporarily look $2+2$.
The cardinality is reproduced as
$|\mathcal{P}_s(m)|  = \frac{9}{9-8}
\binom{5}{1}\binom{3}{0}\binom{1}{1}=45$.
\end{example}

\begin{example}\label{ex:ff}
Consider Example \ref{ex:3s}, which corresponds to $L=19$ and 
$m=(m_1,m_2,m_3,\ldots) = (2,2,1,0,0,\ldots)$,
hence $(p_1,p_2,p_3,\ldots) = (9,3,1,1,1\ldots)$.
Thus (\ref{ff}) with any $s \ge 3$ gives
\begin{align}
|\mathcal{P}_s(m)|  = \frac{19}{1}
\binom{10}{2}\binom{4}{2}\binom{1}{1} = 10260.
\end{align}
\end{example}

\subsection{Generalized Gibbs ensemble}
We are going to study the gas of 
BBS solitons with amplitude not exceeding $s$ 
in terms of the generalized Gibbs ensemble 
involving the energies $E_1, E_2, \ldots, E_s$ with the 
conjugate inverse temperatures 
$\beta_1, \beta_2, \ldots, \beta_s$.
It will be referred to as GGE$(\beta_1, \beta_2, \ldots, \beta_s)$.
The canonical partition function reads
\begin{align}\label{zl}
Z_L(\beta_1,\ldots, \beta_s) = \sum_{\eta}
\mathrm{e}^{-\beta_1 E_1-\cdots - \beta_s E_s}
\end{align}
with the sum taken over $\eta \in \{0,1\}^L$ such that 
$m_j=0$ for $j >s$.
The free energy per site is 
\begin{align}
\mathcal{F} &= -\lim_{L \rightarrow \infty}\frac{1}{L}\log 
Z_L(\beta_1,\ldots, \beta_s)
\label{fz}
\\
&= -\lim_{L \rightarrow \infty}\frac{1}{L}\log \left(
\sum_{m}|\mathcal{P}_s(m)| 
\mathrm{e}^{-\sum_{1 \le j, k \le s}\beta_j\min(j,k)m_k}\right),
\label{fp}
\end{align}
where the sum is taken with respect to 
$m=(m_1, \ldots, m_s) \in (\Z_{\ge 0})^s$.
To get the second line, we used the fact that the inclusion of 
the contributions from those $\eta$ such that $M \ge L/2$
in (\ref{zl}) at most doubles the quantity in $\log$ by 
Remark \ref{re:01}.
In \ref{app:tr}, we present a transfer matrix formalism 
of the partition function (\ref{zl}), which  
formally corresponds to the maximal choice $s=[L/2]$.
It enables us to compute the free energy $\mathcal{F}$.
However to compute the current, we actually need a more refined information on the density 
on solitons with each amplitude. 
(See \ref{app:current2} for an approach to bypass it at least for the
two temperature GGE, although.)
To extract it, we resort to the thermodynamic Bethe ansatz in the next subsection. 

\subsection{Thermodynamic Bethe ansatz}
Assuming the $L$-linear scaling 
\begin{align}
&m_j \simeq L \rho_j,\qquad
p_j \simeq L \sigma_j, \qquad 
E_j \simeq L \varepsilon_j,
\label{scal}
\\
&\varepsilon_j = \sum_{k = 1}^s\min(j,k)\rho_k,
\qquad \sigma_j=1-2\varepsilon_j,
\label{epsi}
\end{align}
and substituting (\ref{ff}) into (\ref{fp}), we find 
\begin{align}\label{Fdef}
\mathcal{F} = \beta_1\varepsilon_1 + \cdots + 
\beta_s \varepsilon_s -\sum_{i=1}^s
\bigl((\sigma_i+\rho_i)\log(\sigma_i+\rho_i)
-\sigma_i \log \sigma_i - \rho_i\log \rho_i\bigr),
\end{align}
where $\rho_i$ is the solution to the condition
$\frac{\partial {\mathcal F}}{\partial \rho_i}=0$.
It leads to the TBA equation:
\begin{align}\label{tba}
\sum_{j=1}^s\min(i,j)\beta_j 
= \log(1+Y_i) - 2\sum_{j=1}^s\min(i,j) \log(1+Y^{-1}_j),
\quad Y_i = \frac{\sigma_i}{\rho_i}.
\end{align} 
It is cast into the (constant) Y-system:
\begin{align}
Y_1^2 &= \mathrm{e}^{\beta_1}(1+ Y_2), 
\label{y1}
\\
Y_i^2 &= \mathrm{e}^{\beta_i}(1+ Y_{i-1})(1+Y_{i+1})\quad (1 < i < s),
\label{sw}
\\
Y_s^2 & = \mathrm{e}^{\beta_s}(1+Y_{s-1})(1+Y_s).
\label{ys}
\end{align}
Note the irregularity of the very last factor which is {\em not} $1+Y_{s+1}$.
In terms of the positive solution $Y_1,\ldots, Y_s$ to the $Y$-system, 
the free energy (\ref{fz}) is given by 
\begin{align}\label{efe}
\mathcal{F} = -\sum_{i=1}^s \log(1+ Y^{-1}_i).
\end{align}
In \ref{app:A} 
we present low temperature series expansions for $\mathcal{F}, Y_i, \rho_i$ and $\varepsilon_i$, etc.

\subsection{Two temperature GGE}\label{ss:2gge}
In this subsection we focus on the special case 
GGE$(\beta_1, \beta_\infty)$ 
including the two inverse temperatures $\beta_1$ and $\beta_\infty$.
Their conjugate energies $E_1$ and $E_\infty$ 
are the only cases that become sums of 
{\em local} correlations. See (\ref{e1}) and (\ref{ei}).
 
Consider the Y-system (\ref{y1})--(\ref{ys})  with 
$\beta_2 = \cdots = \beta_{s-1}=0$.
It is equivalent to the following equations and the boundary condition:
 \begin{align}
&Y_i^2 = (1+ Y_{i-1})(1+Y_{i+1})\;\;(1 \le i \le s),
\label{yy1}
\\
&1+Y_0 = \mathrm{e}^{\beta_1},\quad 
1+Y_{s+1} = \mathrm{e}^{\beta_s}(1+Y_s).
\label{yy2}
\end{align}
An advantage of (\ref{yy1}) over (\ref{sw}) 
is that it allows a general solution simply expressible in terms 
of two parameters $a, z$: 
\begin{align}
Y_i = Q_{i-1}Q_{i+1},\qquad 
Q_i = \frac{a^{\frac{1}{2}}z^{\frac{i}{2}}
-a^{-\frac{1}{2}}z^{-\frac{i}{2}}}
{z^{\frac{1}{2}}-z^{-\frac{1}{2}}},
\end{align}
where the latter satisfies another difference equation called Q-system:
\begin{align}
Q_i^2 = Q_{i-1}Q_{i+1}+1 \quad (i \in \Z).
\end{align}
Moreover, the boundary condition (\ref{yy2}) relates 
$a, z$ to the temperatures as
\begin{align}\label{qlim}
\mathrm{e}^{\frac{\beta_1}{2}} = Q_0 
&= \frac{a^{\frac{1}{2}}-a^{-\frac{1}{2}}}
{z^{\frac{1}{2}}-z^{-\frac{1}{2}}},
\qquad
\mathrm{e}^{\beta_s} = 
\Bigl(\frac{Q_{s+1}}{Q_s}\Bigr)^{\!2}
\;\;\overset{s \rightarrow \infty}{\longrightarrow}\;\;
\mathrm{e}^{\beta_\infty} = z^{-1},
\end{align}
where we have assumed $0 < z < 1$ without loss of generality
to derive the latter relation.
On the other hand the first relation demands $0 < a \le z$ in order 
to satisfy $\beta_1 \ge 0$.

The free energy (\ref{efe}) is evaluated as
\begin{align}
\mathcal{F} &= -\sum_{i=1}^s\log\left(\frac{Q_i^2}{Q_{i-1}Q_{i+1}}\right)
= \log\left(\frac{Q_0Q_{s+1}}{Q_1Q_s}\right)
\;\;\overset{s \rightarrow \infty}{\longrightarrow}\;\;
\log\left(\frac{1-a}{1-az}\right).
\label{tatsuki}
\end{align} 
From ${\mathcal F}$,  the expectation values of the energy densities $\varepsilon_1$ 
and $\varepsilon_\infty$ are derived as 
\begin{align}
\varepsilon_1 &= \frac{\partial {\mathcal F}}{\partial \beta_1}= \left(
\frac{\partial a}{\partial \beta_1}\frac{\partial }{\partial a}
+\frac{\partial z}{\partial \beta_1}\frac{\partial }{\partial z}\right)
\mathcal{F}
= -\frac{a(1-a)}{1+a}\frac{\partial}{\partial a}{\mathcal F}
= \frac{a(1-z)}{(1+a)(1-az)},
\label{ee1}
\\
\varepsilon_\infty &=
\frac{\partial {\mathcal F}}{\partial \beta_\infty}= \left(
\frac{\partial a}{\partial \beta_\infty}\frac{\partial }{\partial a}
+\frac{\partial z}{\partial \beta_\infty}\frac{\partial }{\partial z}\right)
\mathcal{F}
= \left(-
\frac{a(1-a)(1+z)}{(1+a)(1-z)}\frac{\partial }{\partial a}
-z\frac{\partial }{\partial z}\right)
\mathcal{F}=
\frac{a}{1+a},
\label{ee2}
\end{align}
where the latter is the total density of balls due to (\ref{ei}).
Note that $\varepsilon_\infty$  is {\em not} $\frac{z}{1+z}$
despite that $z= \mathrm{e}^{-\beta_\infty}$  from (\ref{qlim}).

So far, we have solved the Y-system (\ref{yy1})--(\ref{yy2}).
The remaining task is to find $\varepsilon_j, \rho_j, \sigma_j$ satisfying 
(\ref{epsi}), $\sigma_i/\rho_i=Y_i$ (\ref{tba}) 
and the boundary condition  (\ref{ee1})--(\ref{ee2}) 
in the limit $s\rightarrow \infty$.
The answer is given by
\begin{align}
\varepsilon_i &= \frac{a(1-z^i)}{(1+a)(1-az^i)},
\label{epaz}
\\
\sigma_i &=  \frac{(1-a)(1+a z^i)}{(1+a)(1-az^i)},
\label{sigaz}
\\
\rho_i &= \frac{az^{i-1}(1-a)(1-z)^2(1+az^i)}
{(1+a)(1-az^{i-1})(1-az^i)(1-az^{i+1})}. 
\label{roaz}
\end{align}
The result  (\ref{sigaz}) reduces to $\varphi^{(1)}_1|_{\kappa=1}$ in 
\cite[eq.(182)]{KL20} when $a=z=q$.
In view of (\ref{qlim}), it corresponds to $\beta_1=0$ which is the 
Gibbs ensemble with the single inverse temperature $\beta_\infty$.
This is the unique case in which the probability
distribution of the BBS local states $\eta_i$  
becomes i.i.d. in which the relative probability is given by 
$\Bbb{P}(\eta_i = 1)/\Bbb{P}(\eta_i=0)=z$.

%--------------------------------------------------
\section{Stationary current}\label{sec:current}
%--------------------------------------------------

%--------------------------------------------------
\subsection{Effective speed of solitons}
%--------------------------------------------------

Consider the BBS with time evolution $T_l$ as a gas of solitons.
Each soliton is subject to the interaction with the others via collisions.
As explained along Example \ref{ex:1s} and Example \ref{ex:2s},
an $i$-soliton has the bare speed $\min(i,l)$ and acquires 
the phase shift $2\min(i,k)$ in its trajectory by a collision with a $k$-soliton.
Let $v^{(l)}_i$ denote the effective speed of $i$-solitons under $T_l$.
Then the above features of the soliton interaction 
lead to the consistency condition \cite{Zakharov_1971,El_2003,FNRW}:
\begin{align}\label{spv}
v^{(l)}_i = \min(i,l) 
+ 2\sum_{k=1}^\infty\min(i,k)(v^{(l)}_i-v^{(l)}_k)\rho_k\qquad (\forall i \ge 1),
\end{align} 
where $\rho_k$ is the density of $k$-solitons.
The first term on the RHS is the bare speed and the 
second term takes the interaction effect into account.
A speed equation of this kind has been postulated in the 
generalized hydrodynamics \cite[eq.(7)]{DYC18}.
In particular (\ref{spv}) reduces to \cite[eq.(11.7)]{FNRW}
as $l \rightarrow \infty$. 

By the definition, the current $J^{(l)}$ in our BBS 
is the number of balls passing through a site to the right by
applying $T_l$ once.
It consists of contributions from $i$-solitons for any $i\ge 1$ as
\begin{align}\label{Jl}
J^{(l)} = \sum_{i=1}^\infty i \rho_i v^{(l)}_i,
\end{align}
where the effective speed $v^{(l)}_i$ should be determined from 
(\ref{spv}) for a given density distribution $(\rho_k)_{k \ge 1}$.
This formula corresponds to \cite[eq.(3.62)]{D19}.

%--------------------------------------------------
\subsection{Coincidence with time average}
%--------------------------------------------------

As demonstrated in Example \ref{ex:3s}, 
balls in a BBS state are moved to the right periodically by a carrier of capacity $l$ 
under the time evolution $T_l$.
Thus the stationary current $J^{(l)}$ (\ref{Jl}) based on the soliton picture 
should coincide with the 
{\em time average} $\bar{J}^{(l)}$ of the number of balls in the carrier over the 
{\em Poincar\'e cycle} of a given BBS state.
An explicit formula of $\bar{J}^{(l)}$ has been obtained in \cite{KT10}
by a calculation involving a tropical 
(or ultradiscrete) analogue of the Riemann theta function. 
In this subsection we prove 
\begin{align}\label{JJ}
J^{(l)} = \bar{J}^{(l)}
\end{align}
quite generally without recourse to a specific choice of 
the soliton densities $(\rho_k)_{k \ge 1}$.
By so doing we will illuminate an intriguing connection between 
the effective speed and the period matrix of the 
tropical Riemann theta function \cite[eq.(4.14)]{KT10}. 
The latter has emerged from the Bethe ansatz at $q=0$.
Its application to the BBS has revealed 
the Bethe roots at $q=0$ as action-angle variables, 
Bethe eigenvalues at $q=0$ giving the Poincar\'e cycle, 
torus decomposition interpretation of a Fermionic character formula, 
multiplicity formula of the invariant torus and 
explicit solution of the initial value problem in terms of 
tropical Riemann theta functions and so forth \cite{KTT06,KT10}.

\subsubsection{Time averaged current and topical period matrix}
The time average $\bar{J}^{(l)}$ for a given state $\eta$ 
has been obtained in \cite[Prop.4.3]{KT10}.
It is independent of the position and expressed in terms of the data 
determined from the conserved Young diagram $Y(\eta)$ (\ref{Yeta}) as
\begin{align}\label{jb}
\bar{J}^{(l)} = \hat{\kappa}^{(\infty){\rm T}} B^{-1}\hat{\kappa}^{(l)}.
\end{align} 
This was the first nontrivial dynamical characteristic of the BBS derived 
from an explicit calculation using the tropical Riemann theta function.
To explain the RHS of (\ref{jb}), let $m_i$ be the number of $i$-solitons and 
assume the same notations for the vacancy  
$p_i$ (\ref{pj}) and the densities $\rho_i, \sigma_i$  (\ref{scal}) as before.
Let $s$ be the width of the Young diagram 
$Y(\eta)$ (\ref{Yeta}), i.e., the amplitude of the largest solitons.
Denote the depth of $Y(\eta)$ by $g= E_1 = m_1+\cdots + m_s$.
Thus in particular the vacancy (\ref{pj}) reads 
\begin{align}\label{pjs}
p_j = L - 2\sum_{k =1}^s\min(j,k)m_k \quad (j \ge 0),
\end{align}
which satisfies $L=p_0 > p_1> \cdots > p_s = \cdots = p_\infty = L-2M \ge 1$.
Now $B$ and $\hat{\kappa}^{(l)}$ are a $g$-dimensional matrix and a column vector 
possessing the block structure as follows \cite[eq.(4.27)]{KT10}:
\begin{align}
B &= \begin{pmatrix}
B_{11}  & \cdots & B_{1s} \\
\vdots  & \ddots  & \vdots \\
B_{s1} & \cdots & B_{ss}
\end{pmatrix},
\quad
B_{ij}= \bigl(\delta_{ij}\delta_{\alpha \beta}p_i 
+2\min(i,j)\bigr)_{1\le \alpha \le m_i, 1 \le \beta \le m_j}
\in \mathrm{Mat}(m_i,m_j),
\label{bmat}
\\
\hat{\kappa}^{(l)}&= (\kappa^{(l)}_i)_{1 \le i \le s, 1\le \alpha \le m_i} \in (\Z_{>0})^g,\qquad 
 \kappa^{(l)}_i = \min(i,l).
 \label{kapa}
\end{align}
The matrix $B$ is the tropical analogue 
of the period matrix of the Riemann theta function
and $\hat{\kappa}^{(l)}$ is the velocity vector of the angle variables in the Jacobi variety
on which the dynamics becomes a straight 
motion \cite{KTT06, KT10}.\footnote{
It was denoted by ${\bf h}_l$ in \cite[eq.(4.27)]{KT10}.}
The vector $\hat{\kappa}^{(\infty){\rm T}}$ 
in (\ref{jb}) denotes the transpose of the column vector $\hat{\kappa}^{(\infty)} $.

\begin{example}
Consider the case $m_1=m_2=m_3=2\,(s=3, g=6)$.
Then $B$ \cite[eq.(4.6)]{KT10} is a $6 \times 6$ matrix and 
$\hat{\kappa}^{(l)}$ is a $6$-dimensional row vector given by
\begin{align}\label{Be}
B&= \begin{pmatrix}
p_1+ 2 & 2 & 2 & 2 & 2 & 2 
\\
2 & p_1+ 2 & 2 & 2 & 2 & 2
\\
2 & 2 & p_2 + 4 & 4 & 4 & 4
\\
 2 & 2 & 4 & p_2 + 4 & 4 & 4 
\\
2 & 2 & 4 & 4 & p_3+6 & 6
\\
2 & 2 & 4 & 4 & 6 & p_3+6
\end{pmatrix},
\\
\hat{\kappa}^{(l){\rm T}}&= \bigl(\min(l,1),\min(l,1),\min(l,2),
\min(l,2),\min(l,3),\min(l,3)\bigr).
\end{align}
\end{example}

\begin{example}
For Example \ref{ex:3s} and Example \ref{ex:ff}, we have
\begin{align}
B=\begin{pmatrix}
11 & 2 & 2 & 2 & 2
\\
2 & 11 & 2 & 2 & 2
\\
2 & 2 & 7 & 4 & 4
\\
2 & 2 & 4 & 7 & 4
\\
2 & 2 & 4 & 4 & 7
\end{pmatrix},
\quad
\hat{\kappa}^{(1)} = \begin{pmatrix} 1 \\ 1 \\ 1 \\ 1 \\ 1 \end{pmatrix},
\quad
\hat{\kappa}^{(2)} = \begin{pmatrix} 1 \\ 1 \\ 2 \\ 2 \\ 2 \end{pmatrix},
\quad
\hat{\kappa}^{(l)} = \begin{pmatrix} 1 \\ 1 \\ 2 \\ 2 \\ 3 \end{pmatrix}\;\; (l\ge 3).
\end{align}
\end{example}

\subsubsection{Inverse of tropical period matrix}
To elucidate the connection to the speed equation, our first task is to compute the inverse $X= B^{-1}$.
It turns out that $X$ also has the same block structure as $B$:
\begin{align}
X &= \begin{pmatrix}
X_{11}  & \cdots & X_{1s} \\
\vdots  & \ddots  & \vdots \\
X_{s1} & \cdots & X_{ss}
\end{pmatrix},
\quad
X_{ij}= \bigl(\delta_{ij}\delta_{\alpha \beta}p_i^{-1} 
+x_{\min(i,j)}\bigr)_{1\le \alpha \le m_i, 1 \le \beta \le m_j}
\in \mathrm{Mat}(m_i,m_j).
\end{align}
The condition$\sum_{j=1}^sB_{ij}X_{jk} = \delta_{ik}\mathrm{id}_{m_i}$
reads, in terms of the parameters $x_1, \ldots, x_s$, as
\begin{align}\label{ta}
p_ix_{\min(i,j)} + 2\min(i,j)p_j^{-1} +2\sum_{k=1}^s \min(i,k)m_k x_{\min(j,k)}=0
\quad (1 \le i,j \le s).
\end{align} 
For instance when $s=4$, it can be presented in a matrix form:
\begin{align}\label{BX}
{\footnotesize
\begin{pmatrix}
p_1+2m_1 & 2m_2  & 2m_3 & 2m_4 
\\
2m_1 & p_2+4m_2  & 4m_3 & 4m_4 
\\
2m_1 & 4m_2 & p_3+6m_3 & 6m_4
\\
2m_1 & 4m_2 & 6m_3 & p_4+ 8m_4
\end{pmatrix}
\begin{pmatrix}
x_1 & x_1 & x_1 & x_1
\\
x_1 & x_2 & x_2 & x_2
\\
x_1 & x_2 & x_3 & x_3
\\
x_1 & x_2 & x_3 & x_4
\end{pmatrix} = -2 
\begin{pmatrix}
p_1^{-1} & p_2^{-1} & p_3^{-1} & p_4^{-1}
\\
p_1^{-1} & 2p_2^{-1} & 2p_3^{-1} & 2p_4^{-1}
\\
p_1^{-1} & 2p_2^{-1} & 3p_3^{-1} & 3p_4^{-1}
\\
p_1^{-1} & 2p_2^{-1} & 3p_3^{-1} & 4p_4^{-1}
\end{pmatrix}.}
\end{align}
In \ref{app:note29}, we prove that the over-determined $s^2$ equations (\ref{ta})  on 
$x_1,\ldots, x_s$ admits the unique solution:
\begin{align}\label{kawa}
x_k = -\sum_{j=1}^k\frac{2}{p_{j-1}p_j}.
\end{align}

\subsubsection{Speed equation as inversion of tropical period matrix}
By substituting (\ref{pjs}) into the first term of (\ref{ta}) 
with $j$ replaced by $l$,  we have
\begin{align}
Lx_{\min(i,l)}&= 
-\frac{2\min(i,l)}{p_l}
+2\sum_{k=1}^s\min(i,k)\bigl(
x_{\min(i,l)}-x_{\min(l,k)}\bigr)m_k.
\end{align}
Further substituting the scaling forms (\ref{scal}) and 
\begin{align}\label{xnu}
x_k \simeq L^{-2} \nu_k
\end{align}
and taking the limit $L, s \rightarrow \infty$ we get
\begin{align}
\nu_{\min(i,l)}&= 
-\frac{2\min(i,l)}{\sigma_l}
+2\sum_{k=1}^\infty\min(i,k)\bigl(
\nu_{\min(i,l)}-\nu_{\min(l,k)}\bigr)\rho_k
\quad (\forall i, l \ge 1).
\end{align}
Comparing this with (\ref{spv}), 
we find that the speed $v^{(l)}_i$ of
$i$-solitons under $T_l$ is given by 
\begin{align}\label{vnu}
v^{(l)}_i = -\frac{\sigma_l}{2}\nu_{\min(i,l)}
= \frac{\sigma_l}{\sigma_\infty} v_{\min(i,l)},
\qquad 
v_i := v^{(\infty)}_i = -\frac{\sigma_\infty}{2}\nu_i.
\end{align}
Thus the effective speed of solitons can essentially be identified with 
appropriately scaled elements of the inverse tropical period matrix.
Our subsequent proof of (\ref{jb}) in Sec. \ref{ss:proof} will further demonstrate this fact 
somewhat more directly.

Combining (\ref{kawa}), (\ref{xnu}) and (\ref{vnu}) 
we discover the general relation connecting the effective speed 
and the hole density  ($\sigma_0 = 1$):
\begin{align}\label{vsig}
v^{(l)}_k = \sum_{j=1}^{\min(k,l)}
\frac{\sigma_l}{\sigma_{j-1}\sigma_j},
\qquad
v_k = \sum_{j=1}^k\frac{\sigma_{\infty}}{\sigma_{j-1}\sigma_j}.
\end{align}
We note that these are formally exact even for {\em finite} $L$ 
provided that all the $\simeq$ in (\ref{scal}) and (\ref{xnu}) are replaced by the equality.

In \ref{app:current1}, we evaluate this sum 
for (\ref{sigaz}) to obtain the effective speed and the current explicitly 
in GGE$(\beta_1, \beta_\infty)$.
The result are (\ref{mirei}) and (\ref{chizuko}).
In \ref{app:current2}, we further confirm the result 
by a transfer matrix method.

\subsubsection{Proof of (\ref{JJ})}\label{ss:proof}

Given a column vector  $\xi = (\xi_i) = (\xi_1, \xi_2,\ldots)^{\rm T}$ 
whose components are labeled by soliton amplitude $i$, we denote its $m_i$-fold duplication for each $i$ by
attaching hat as 
\begin{align}
\hat{\xi} = (\overset{m_1}{\overbrace{\xi_1, ..., \xi_1}}, \overset{m_2}{\overbrace{\xi_2, ... , \xi_2}}, \ldots)^{\rm T}.
\end{align}
This convention matches the notation in (\ref{kapa}). 
Note that the speed equation (\ref{spv})  is expressed by using the vacancy $p_i = L - 2\sum_j \min(i,j) m_j$ as
\begin{align}\label{sp1}
p_i v^{(l)}_i + \sum_j 2\min(i,j) v^{(l)}_j m_j = L \kappa^{(l)}_i.
\end{align}
Recalling that $B= (\delta_{ij} \delta_{\alpha \beta} p_i + 2\min(i,j))_{i\alpha, j\beta}$ (\ref{bmat}), 
the equation (\ref{sp1}) is written in the matrix form
\begin{align}
B \hat{v}^{(l)} = L \hat{\kappa}^{(l)}.
\label{disguised}
\end{align}
On the other hand the soliton current density is
\begin{align}
J^{(l)} = \sum_i i\rho_i v^{(l)}_i = \frac{1}{L}\sum_i im_i v^{(l)}_i = 
\frac{1}{L} \hat{\kappa}^{(\infty) {\rm T}}\cdot \hat{v}^{(l)}.
\end{align}
Thus we have the equality with the time average as
\begin{align}
J^{(l)} = \frac{1}{L} \hat{\kappa}^{(\infty) {\rm T}}\cdot L B^{-1} \hat{\kappa}^{(l)}
=
\hat{\kappa}^{(\infty) {\rm T}}\cdot B^{-1}\hat{\kappa}^{(l)} = \bar{J}^{(l)}.
\end{align}

It will be interesting to extend the result of this subsection to the 
periodic box-ball system with $n$ kinds of balls \cite{KT10}.
In fact, most of the necessary ingredients have already been formulated universally 
in terms of the Bethe ansatz and the root system associated with $A^{(1)}_n$ 
including the time averaged current  \cite[Prop.~4.4]{KT10}
and the tropical period matrix \cite[eq.(4.6)]{KT10}.

%--------------------------------------------------
\section{Domain wall problem and generalized hydrodynamics}\label{sec:5}
%--------------------------------------------------

In this section we discuss the dynamics of the BBS model when the system is initially prepared
in an inhomogeneous state. 
We  consider in particular a family of domain-wall problems with different ball densities in the ``left'' and in the ``right'' halves of the system.

To be precise, at time $t=0$, in the ``left'' half of the system (from site $r=-L/2$ to $r=0$), we have, on each site, a ball with probability $p_L$
and an empty site with probability $1-p_L$. In term of fugacity, we have  $z_L: p_L=z_L/(1+z_L)$.
Similarly, in the ``right'' half (from site $r=1$ to $r=L/2-1$) the ball density is $p_R$. So, each halve of the system is thus initially prepared in a one-temperature GGE state.
Next we evolve, using $T_l$ with carrier capacity $l$, and mainly investigate the evolution
of the ball density $h^{(l)}(r,t)$ at site $r$ and time $t$,
averaged over a large number of initial states.

We present data for $l=2,3,4,10,20$ and 100, and the simulations we carried out up to $t=2000$.
Unless specified otherwise the simulations have been performed with $N_{\rm samples}=5.10^4$ random initial conditions and a system size $L=10^5$. In a few cases, to increase the precision, we pushed the simulations to  $N_{\rm samples}=10^6$ and $L=10^6$.
The systems we simulate are sufficiently large, so that, in the time range we consider the ``wrapping effect'' on the periodic ring does not play any role and we are therefore effectively dealing with the {\em infinite} system. 

\begin{figure}[h]
\begin{center}
\includegraphics[width=0.65\textwidth]{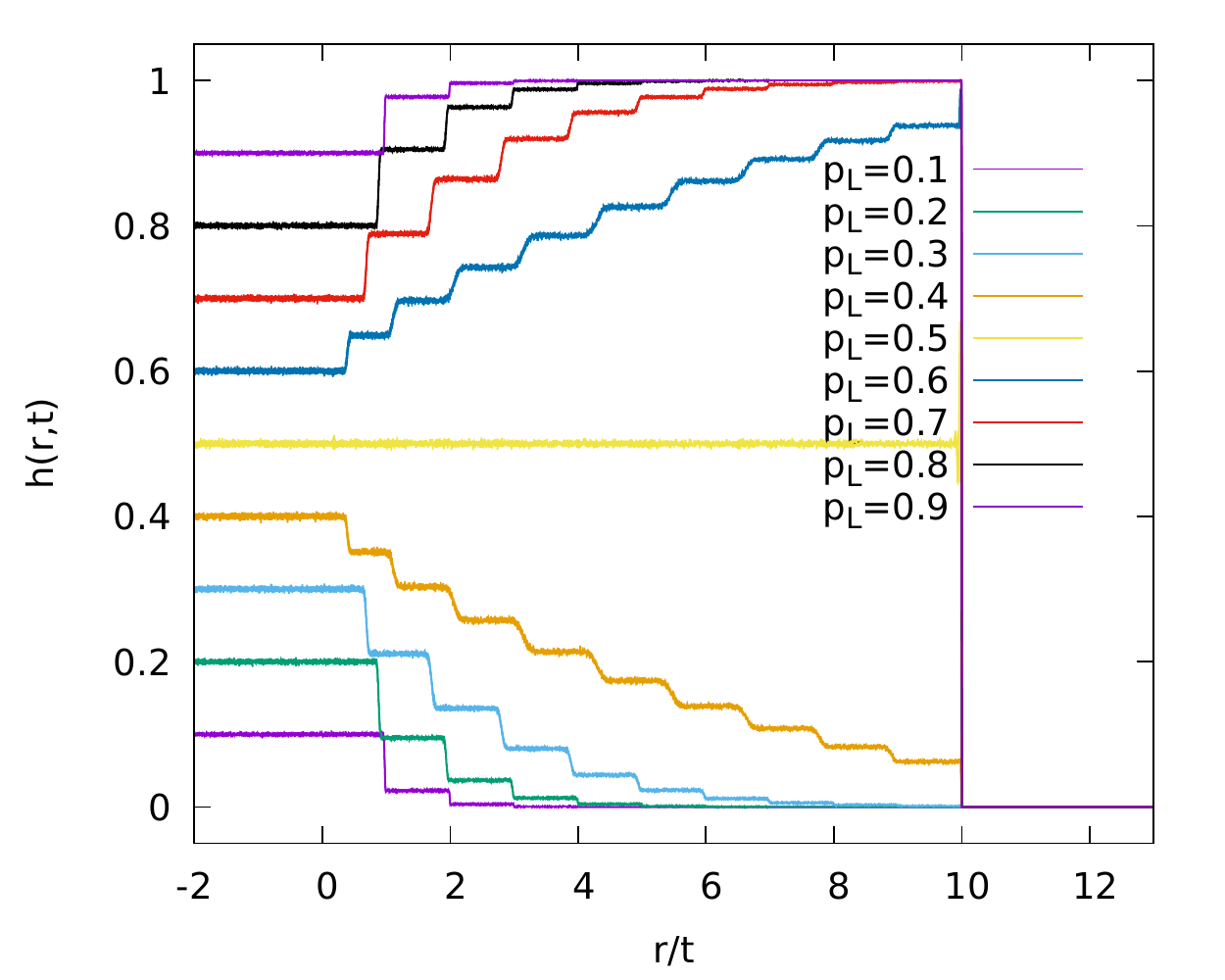}
\caption{Ball density plotted as a function $\zeta=r/t$ for different initial ball densities $p_L$ in the left half of the system. The right half is initially empty, $p_R=0$, and $l=10$. 
For each density $p$ two curves are plotted, corresponding to $t=500$ and $t=1000$. Both curves turn out to be almost on top of each other. 
The density vanishes for $\zeta>l=10$, which corresponds to the fastest velocity in this problem. Note that for $\zeta<l$ the curves associated to $p$ and $1-p$ are symmetric
with respect to $h=1/2$ (see Remark \ref{re:01}).
} \label{Fig:l=10_pR=0}
\end{center}\end{figure}

The figure \ref{Fig:l=10_pR=0} represents the ball density $h^{(l)}(r,t)$
for an initial state with densities $p_R=0$, various values of $p_L$, and $l=10$.
When plotted as a function of $\zeta=r/t$, the data associated
to different times $t$ practically collapse onto a single curve and these density profiles show some marked plateaux. As we will discuss in the following paragraphs, this can be understood and described analytically in terms of
generalized hydrodynamics (GHD) \cite{Castro-Alvaredo2016,Bertini2016,doyon_dynamics_2017,DYC18,D19}
(see also \cite{el_kinetic_2005,kamchatnov_kinetic_2011})
adapted to the BBS. We also note that a similar domain-wall problem for a classical integrable model of hard-rods has been recently solved using GHD \cite{doyon_dynamics_2017}.

The central idea is to assume that the system can locally, at some Euler scale, be described by 
soliton densities $\rho_j$ and effective velocities $v_j$ related through (\ref{vitesse0}). The evolution in space and time of these quantities is then obtained by imposing
the conservation of each soliton species (see (\ref{eq:jkrhokvk})  and (\ref{cons})).

%--------------------------------------------------
\subsection{Densities and velocities}
%--------------------------------------------------

We first introduce some notations:  if $v$ is a vector, each entry $v_k$ being associated to one soliton size, we denote ${\bf v}_{kl}=v_k\delta_{kl}$ the corresponding diagonal matrix. If $v,w$ are vectors we denote by $v*w$ the vector with components $(v*w)_k=v_k w_k$. It is also equal to $v*w={\bf v}w=w*v$.
Also, ${\bf v}^{-1}w=w/v$. Note that for a matrix $M$, in general  $(Ma)*b\ne M(a*b)$ unless $M$ is diagonal.

In order to rewrite the equation (\ref{spv}) for the velocities,  
we define the scattering shift matrix\footnote{The diagonal of $M$ does not enter (\ref{vitesse0}), and this leads to several possible choices for the matrix $M$.}
\begin{equation}
M_{kj}=2 \min(k,j).
\label{Mdef}
\end{equation}
For the time evolution generated by $T_l$, the effective velocity $v_k^{(l)}$ of amplitude-$k$ solitons in a state with soliton densities $\left\{\rho_j\right\}$ 
satisfies  (\ref{spv}), i.e.,
\begin{equation}
v_k^{(l)}=\kappa^{(l)}_k-\sum_{j=1}^\infty M_{kj}(v_j^{(l)}-v_k^{(l)})\rho_j.
\label{vitesse0}
\end{equation}
In the expression above the vector $\kappa^{(l)}$ encodes the bare soliton velocities and is defined by  
$\kappa_k^{(l)}=\min(k,l)$ as in (\ref{kapa}).
In this section we will however omit the superscript $(l)$ for all quantities, for brevity. 
We thus have, in a compact form
\begin{equation}
(1-M\rho)*v=\kappa-M(\rho* v).
\label{vitesse}
\end{equation}
Next we know that the hole density is given by  (\ref{epsi}):
\begin{equation}
\sigma=1-M\rho \label{eq:def_sigma}
\end{equation}
with $1$ denoting the vector $(1,1,\cdots)$.
We can define so-called filling fractions $y_i$ by $y_i=\rho_i/\sigma_i$. They correspond to $Y_i^{-1}$ in (\ref{tba}), and
the vector $y$ satisfies
\begin{equation}
y*\sigma=\rho.
\label{eq:def_y}
\end{equation}
The relations (\ref{eq:def_sigma}) and (\ref{eq:def_y}) are equivalent to (\ref{scal}) and (\ref{epsi}).
As we will see, the vector $y$ plays a central role
in the solution of the GHD equations.

The knowledge of $y$ allows us to define a ``dressing'' operation. To any vector $o$ we associate a corresponding
dressed vector $o^{dr}$ which satisfies:
\begin{equation}
o^{dr}
=o-M{\bf y}o^{dr}.
\label{dressed}
\end{equation}
In practice $o^{dr}$ can be computed from $o$ using the inverse of the matrix $1+M{\bf y}$: 
\begin{equation}
o^{dr} =(1+M{\bf y})^{-1}o.
\label{dressed2}
\end{equation}
In particular, (\ref{eq:def_sigma}) and (\ref{eq:def_y}) imply that
\begin{equation}
 \sigma=1-M{\bf y}\sigma \label{eq:rMnr}
\end{equation}
and thus  
\begin{equation}
\sigma=1^{dr}.
\label{rdressed}
\end{equation}
Using (\ref{eq:def_y}) and (\ref{eq:rMnr}) we can rewrite (\ref{vitesse}) as
\begin{equation}
 (1-M{\bf y}\sigma)*v=\kappa-M{\bf y}(\sigma*v)
 \end{equation}
 or
 \begin{equation}
  \sigma*v=\kappa-M{\bf y}(\sigma*v),
  \label{sig*v}
\end{equation}
which means that $\sigma*v=\kappa^{dr}$ and, from (\ref{rdressed})
\begin{equation}
1^{dr}*v=\kappa^{dr}.
\label{vitess2}
\end{equation}
The equation above provides a way to compute the effective velocities from the knowledge of $y$, using the dressing of two known vectors, $1$ and $\kappa$.
This gives naturally rise to a formal expansion in powers of $y$ (but $y$ does not need to be small). Notice that (\ref{eq:rMnr}) coincides with (\ref{pj}), and (\ref{sig*v})  with (\ref{disguised}), in a slightly disguised form. 

\begin{remark}\label{re:M}
Other choices for diagonal part of the matrix $M$ are possible, since
its diagonal does not enter (\ref{vitesse0}).
In particular we can define $n=(1+y^{-1})^{-1}$.
Then, replacing $M$ by $\tilde M=M-1$ modifies the previous expressions as follows: $M\to \tilde M,\ y \to n,\ \sigma \to r=\rho+\sigma$. 
So, (\ref{eq:rMnr}) and (\ref{sig*v}) become:
\begin{align}
r&=1-\tilde{M}{\bf n}r,\\
r*v&=\kappa - \tilde M{\bf n}(r*v),
\label{diagmod}
\end{align}
And the dressing gets modified accordingly.
\end{remark}
For any vector $h$, one can define a charge $q_h=\sum_k h_k\rho_k=h\cdot\rho$ which is conserved. The dressing enables to express $q_h$ in terms of $y$:
\begin{equation}
q_h=h\cdot\rho=h^{dr}\cdot y.
\label{1dressed}
\end{equation}
Indeed:
$$h\cdot\rho=h\cdot{\bf y}1^{dr}=(1+{\bf y}M)^{-1}{\bf y}h\cdot1={\bf y}(1+M{\bf y})^{-1}h\cdot 1=h^{dr}\cdot y$$

The BBS is a concrete realization 
of a {\em linearly degenerate hydrodynamic system} \cite{Tsarev_1991, Bulchandani2017} (see
also \cite{Kamchatnov_nonlinear_2000}), which we briefly describe in \ref{app:degenerate}. 

\subsection{Dynamics of an inhomogeneous system}
We want to consider an inhomogeneous system, with the hypothesis that it
can be locally described using the above formalism, but where all densities have acquired a dependence on space ($x=i/L$, $i=$ lattice index) and time ($t=i/L$, $i$=time step). 
{\bf The main assumption} is that the hole current $(j_{\sigma})_k$ associated to amplitude-$k$ solitons is given by
\begin{equation}
 (j_{\sigma})_k= \sigma_k v_k \label{eq:jkrhokvk}
\end{equation}
and we have the hole conservation equation\footnote{$\sigma_k$ can be replaced by $f_k(y_k)\sigma_k$ and $j_{\sigma}$ by  $f_k(y_k)j_{\sigma}$ with $f_k$ arbitrary. In particular, taking $f_k(y_k)=y_k$, we recover the soliton conservation equation $\partial_t\rho + \partial_x (v *\rho)=0$.\label{foot4}}
\begin{equation}
\partial_t\sigma + \partial_x j_{\sigma}=0.
\label{cons}
\end{equation}
Remarkably, it implies that the curves $y_k$ constant are characteristics. 
Using $\sigma=(1+M{\bf y})^{-1}1$, the above equation rewrites:
\begin{equation}
\partial_t((1+M{\bf y})^{-1}) (1)+\partial_x((1+M{\bf y})^{-1}) (\kappa)=0.
\label{eq:diff}
\end{equation}
One has:
\begin{equation}
\partial_\alpha((1+M{\bf y})^{-1})=-\beta (\partial_\alpha{\bf y}) (1+M{\bf y})^{-1}
\;\;{\rm with}\;\alpha=t\;{\rm or}\;x .
\end{equation}
where $\beta$ is the matrix $\beta=(1+M{\bf y})^{-1}M$.
Factorizing $\beta$ out of  (\ref{eq:diff}) we finally get 
\begin{equation}
 \partial_t y+{\bf v}  \partial_x y=0,
\label{cons1}
\end{equation}
which means that the $y$ are the normal modes of the hydrodynamics \cite{D19}.
We now assume that the system exhibits some ballistic scaling, so that $y_k$ only depends on the rays $\zeta=x/t$.
Then (\ref{cons1})  becomes
\begin{equation}
(\zeta-v_k)\partial_\zeta y_k=0.
\label{consbal}
\end{equation}
The equation above means that $y_k(\zeta)$ must be constant except for possible discontinuities at wave fronts $\zeta=v_k$. 
Since the set of velocities is discrete in this problem, we expect the state of the system
to be piecewise constant in the variable $\zeta$. This will be confirmed by the calculations presented in the following paragraphs, as well as by the simulations.

%----------------------------------------------------------------------------------------------------
\subsection{Current conservation and discontinuities in $y$}
\label{ssec:disc}

We want to show that the filling fraction of a soliton can change discontinuously across the wave front $x/t=v_k$, equal to its speed, without 
violating the current conservation. 
We show it for the following systems which are more general  than  (\ref{eq:rMnr}) and (\ref{sig*v}):
\begin{align}
\sigma&=\alpha-M{\bf y}\sigma, \\
\sigma*v&=\kappa - M{\bf y}(\sigma*v),
\end{align}
where $\alpha$ and $\kappa$ are arbitrary vectors, and $M$ an arbitrary symmetric matrix. 
Let  $y$ denote the filling fraction vector. The dressing is defined as earlier (\ref{dressed}) and  $\sigma=\rho/y=\alpha^{dr}$,  
$v=\kappa^{dr}/\sigma$. 

Let $P=(1+M{\bf y})$, and $D$ its determinant. For $o$ a vector, let us denote $P(o)$ the vector with components $P_i(o)$ equal to the determinants of the matrices obtained by substituting $o$ to the $i$ th column of $P$. 
The dressing can be expressed as: $Do^{dr}=P(o)$. The  $y_k$ dependence of $P$ is only through its $k^{th}$ column. 
With this notation, 
\begin{equation}
v_k={\kappa^{dr}_k\over \alpha^{dr}_k}={P_k(\kappa) \over P_k(\alpha)},
\label{vk-jacobi}
\end{equation}
and as a result, the speed of the soliton $k$
does not depend on $y_k$.

In the frame moving at speed $v_k$, the continuity of the $j$-soliton current takes the form
\begin{equation}
\sigma^L_j(v_j^L-v_k)=\sigma^R_j(v^R_j-v_k)
\label{cotinuity}
\end{equation}
where $L,R$ refer to the  regions $x/t<v_k$ and  $x/t>v_k$. To show that (\ref{cotinuity}) holds when $y^L_k\ne y^R_k$ we need to verify that both sides do not depend on $y_k$. 

Let us denote $P_{ij}(o,o')$ the antisymmetric tensor of the determinants of the matrices obtained by substituting $o$ to the $i$ th column and $o'$ to the $j$ th column of $P$.
The Desnanot-Jacobi identity (also called Sylvester determinant identity) rewrites:
\begin{equation}
P_j(o)P_k(o')-P_k(o)P_j(o')=DP_{jk}(o,o')
\label{jacobi1}
\end{equation}
Taking $o=\kappa$ and $o'=\alpha$ we obtain:
\begin{equation}
\sigma_j(v_j-v_k)={P_{jk}(\kappa,\alpha)\over  P_k(\alpha)}.
\label{cotninuity1}
\end{equation}
Since neither $P_{jk}$ nor  $P_k(\alpha)$
depend on $y_k$, the result follows.\footnote{The same type of argument enables to show that for $M$ given by (\ref{Mdef}), if $\alpha_j=\alpha_l$ and 
$\kappa_j=\kappa_l$ for $j\ge l$, one also has $v_j=v_l$.}

%--------------------------------------------------
\subsection{Domain wall initial condition}\label{ssec:dwghd}
%--------------------------------------------------

Let us solve (\ref{consbal}) for the time evolution of a domain-wall state with  
ball  fugacities  $z_L$ at the left and $z_R$ at the right of the origin.
This initial state defines some occupation vectors $y_L$ and $y_R$
and we have to determine the vectors $y(\zeta)$.
For $\zeta<0$, $y(\zeta)=y_L$, since this region cannot be influenced by the right side.
For $\zeta\gg0$ we have $y(\zeta)=y_R$ since the influence of the left side cannot propagate to the right infinitely fast.

\begin{figure}\begin{center}
\includegraphics[width=0.55\textwidth]{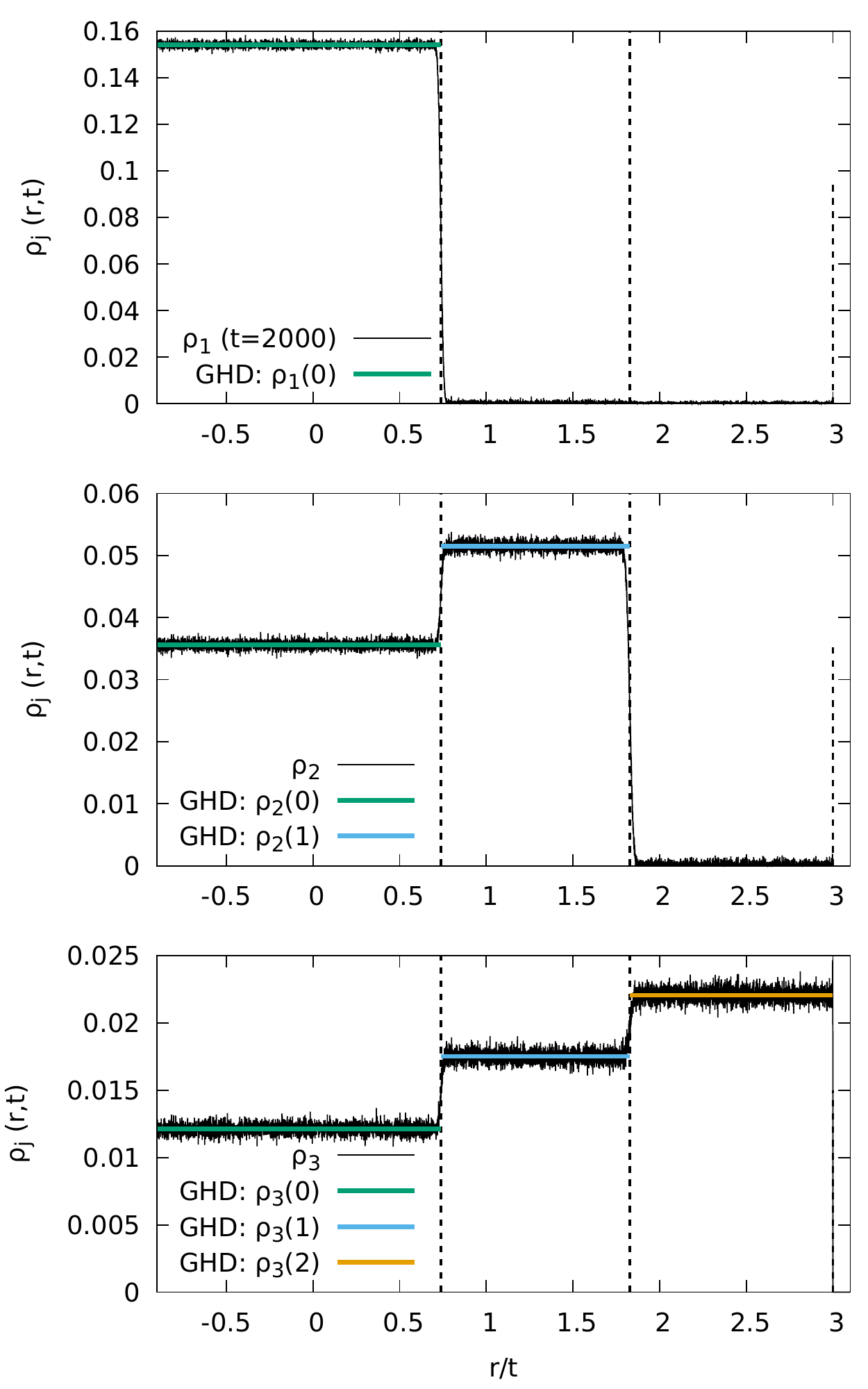}
\caption{Soliton densities $\rho_{i}$ for $i=1$ (top panel), $i=2$ (middle pane) and $i=3$ (bottom).
These densities are plotted as a function $\zeta=r/t$ for $l=3$ and $t=2000$. Here we used $N_{\rm samples}=2.10^5$ random initial conditions
with initial ball densities  $p_L=0.3$  and $p_R=0$ (same parameters as in figure~\ref{Fig:l=3}). One observes the disappearance of the solitons
of size $k$ when going from the plateau $k-1$ to the plateau $k$, as expected from GHD. Furthermore, the densities of the various
types of solitons in each plateau are in quantitative agreement with GHD (the horizontal lines represent 
the value $\rho_i(k)$ at the $k^{\rm th}$ plateau obtained by solving the GHD equations).
}
\label{FigSolitonDensities}
\end{center}\end{figure}

We set $\zeta(k=0)=-\infty$. Then, for each amplitude-$k$ soliton, we can determine the ray $\zeta(k)$ equal to its speed, 
$v_k(\zeta(k))=\zeta(k)$ fixing $y_k(\zeta)=(y_L)_k$ for $\zeta<\zeta(k)$,
$y_k(\zeta)=(y_R)_k$ for $\zeta>\zeta(k) $. The vector  $y(\zeta)$ and therefore, the speeds  $v(\zeta)$ and densities are piecewise constants and 
can jump at the discrete set $\zeta(k)$ coinciding with a $k$-soliton speed. This piecewise structure is illustrated in figure~\ref{FigSolitonDensities},
where $\rho_j(r,t)$ are plotted as a function $\zeta=r/t$ for $k=1,2$ and 3 in a case with $l=3$.
 
Assuming $\zeta(k)$ is an increasing function of $k$, in the sector $k$: $\zeta(k) <\zeta < \zeta(k+1)$, $y_j(\zeta)=(y_R)_j$ for $j\le k$ and $y_j(\zeta)=(y_L)_j$ for $j> k$.
Knowing the occupation vector $y(k)$ in the sector $k$, we know the dressing matrix and we can deduce the associated density vector  $\sigma(k)$ 
(thanks to $\sigma=1^{dr}$ (\ref{rdressed})) and the speed vector $v(k)$ (thanks to $v=\kappa^{dr}/1^{dr}$ (\ref{vitess2})).

Next we determine the boundaries of the $k^{th}$ sector, which are given by the 
speeds of the soliton $k$ and $k+1$ evaluated in this sector $[\zeta(k),\zeta(k+1)]=[v_k(k),v_{k+1}(k)]$.
Note that above we have omitted the $(l)$ superscript, but in general the plateaux boundaries $\zeta(k)$ depend on the capacity $l$ of the carrier
($p_L=0$ is a notable exception, discussed in Sec.~\ref{ssec:pR_only}).

As we have seen in Sec.~\ref{ssec:disc},
the two   determinations of $\zeta(k)$ from the sectors left and right of it coincide, $v_k (k)= v_{k}(k-1)$.
For the $T_l$ evolution,
we can restrict the consideration to 
the $l$ first solitons $k\le l$ since the velocities $v_k$ are all equal for $k\ge l$. So there are $l+1$ sectors $[-\infty,v_1],\cdots,[v_l,\infty]$.  The 
sectors $k>l$  have zero width and their height goes progressively to zero.

\begin{figure}\begin{center}
\includegraphics[width=\textwidth]{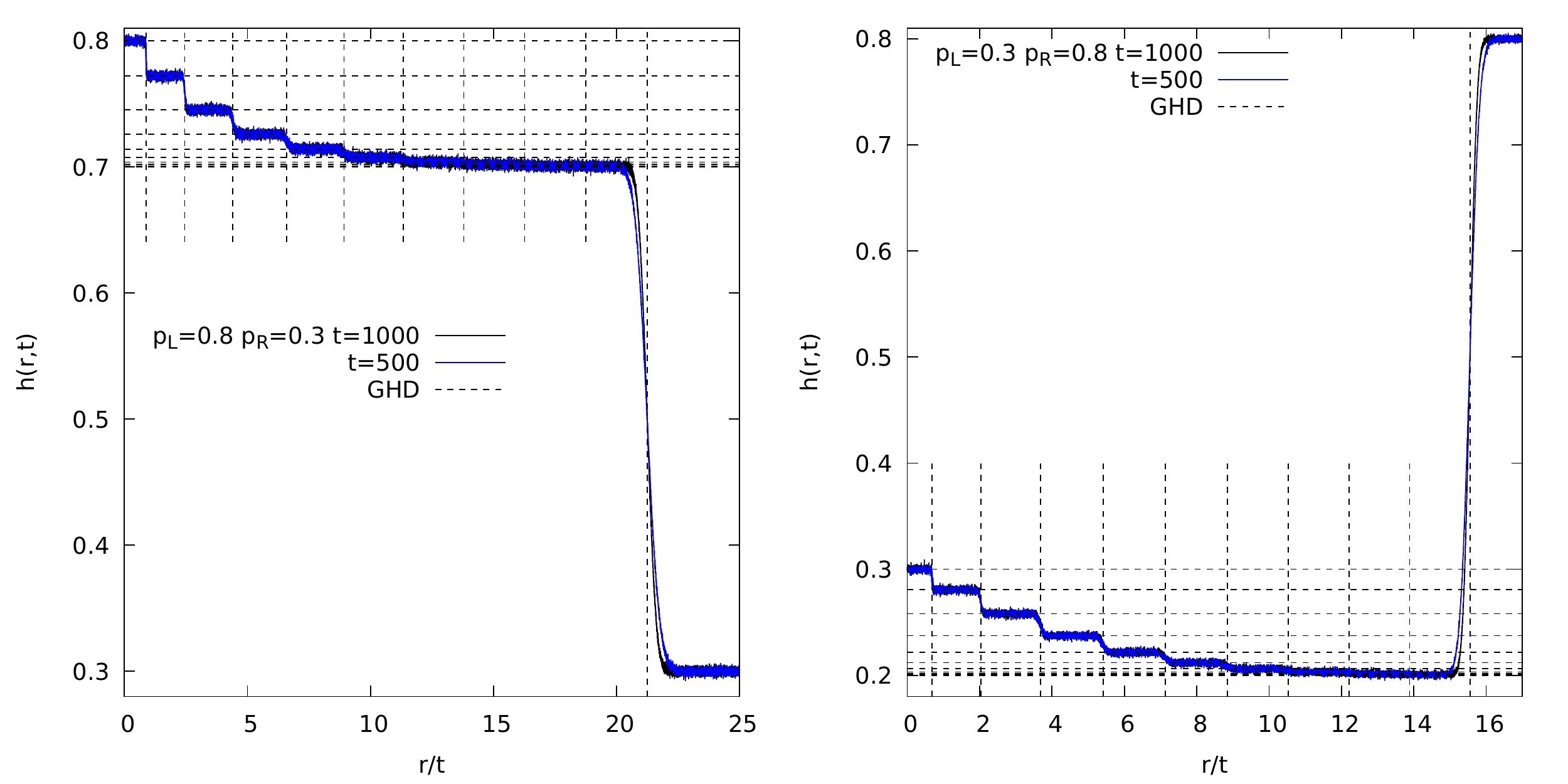}
\caption{Left: ball density for $l=10$, $p_L=0.8$ and $p_R=0.3$. The curves at $t=1000$ (black) and $t=500$ (blue) are practically on top of each other.
Right:  same for $p_L=0.3$ and $p_R=0.8$. Note that in the two cases the long intermediate density plateau turns out to have a ball density equal to $h=1-p_R$ (dotted line).
The dotted lines are the plateau heights and positions obtained by solving numerically the GHD equations (Sec.~\ref{ssec:dwghd}).}
\label{fig:pLpRnz}
\end{center}\end{figure}

The figure~\ref{fig:pLpRnz} presents the ball density measured 
in a simulation for a domain-wall initial state with $p_L>0$ and $p_R>0$.
In the same plot we have shown (dotted lines) the heights and positions of the plateaux  predicted by the method described above, and the agreement with the numerical results is excellent. 

%--------------------------------------------------
\subsection{Case $l=2$}
%--------------------------------------------------

\begin{figure}\begin{center}
\includegraphics[width=\textwidth]{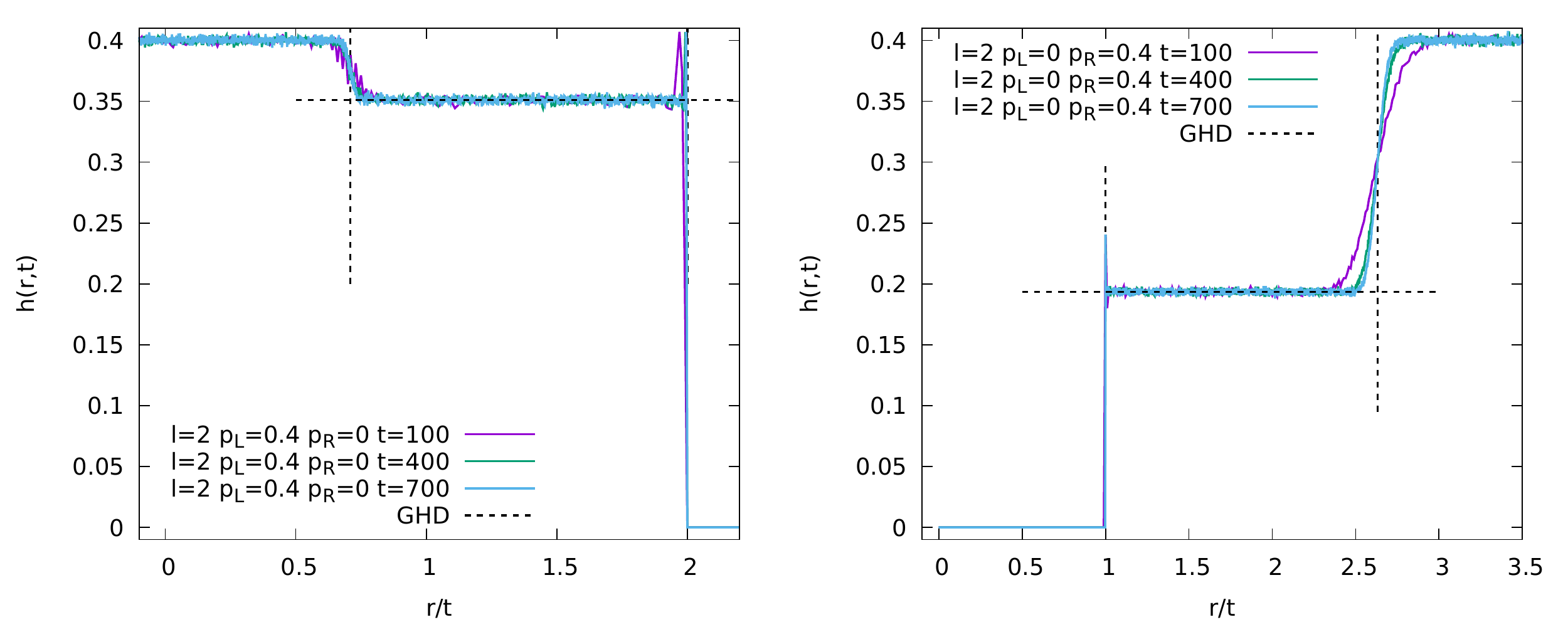}
\caption{Left: Ball density plotted as a function $\zeta=r/t$ for the carrier capacity $l=2$ and different times.
The initial ball densities: $p_L=0.4$ and $p_R=0$.
A zoom on the density spike at $\zeta=2$ is displayed in figure~\ref{FigPeak}.
The horizontal dotted line represents the GHD prediction for the ball density at the nontrivial plateau ($h(1)=0.3511$, see (\ref{eq:h1}) and (\ref{hkL})). The vertical dotted lines
represent the GHD prediction for the end points of this plateau ($\zeta^{(2)}(1)=0.7085$ and $\zeta^{(2)}(2)=2$, see (\ref{eq:zkl})).
Right: same, but with  $p_L=0$ and $p_R=0.4$. In that case the density spike is at $\zeta=1$ and we have $h(1)=0.1935$ (from (\ref{hkR})), $\zeta(1)=1$ and $\zeta(l)=2.6316$ (from (\ref{zetaR})) as solutions of the GHD equations, in good agreement with the simulations.
} \label{fig:l=2}
\end{center}\end{figure}

Equating the current associated to amplitude-$j$ solitons on both sides of the transition from the plateau $k$ and $k+1$
(which is located at $\zeta=\zeta(k+1)=v_{k+1}(k+1)$) reads:
\begin{equation}
 \rho_j(k)\left[v_j(k)- \zeta(k+1)\right] = \rho_j(k+1)\left[v_j(k+1)- \zeta(k+1)\right].
 \label{eq:j_cont_plateaux}
\end{equation}
The equation above was obtained from (\ref{cotinuity}) by replacing $\sigma$ by $\rho$,
which is legitimate thanks to the footnote \ref{foot4}.
In the simple case with $l=2$ these equations can be solved directly, bypassing the use of $y(\zeta)$.
Let us focus for instance on $p_L>0$ and $p_R=0$, as illustrated in the left panel of figure~\ref{fig:l=2}.
The ray space can be divided into three sectors (or plateaux). 
(0) $\zeta<v_1^{(2)}(0)$ where all the speeds $v_j^{(2)}(0)$ and all the densities $\rho_j(0)$ coincide with the homogeneous case.
(1) $v_1^{(2)}(0)<\zeta<v_2^{(2)}(1)=2$ where solitons $j=1$ are absent and all the others move at speed $v^{(2)}_{j\geq2}(1)=2$.
(2) $\zeta>2$ is empty. Across the ray $\zeta(1)=v_1^{(2)}(0)=v_1^{(2)}(1)$,  we use (\ref{eq:j_cont_plateaux}) with $k=0$ to get:
\begin{equation}
 \rho_j(0) \left[v_j^{(2)}(0)- v_1^{(2)}(0)\right] = \rho_j(1)\left[v_2^{(2)}(1)- v_1^{(2)}(0)\right]
\end{equation}
In the sector 1 all solitons with $j\geq2$ move a speed $2$ ($=l$), so the above equation gives:
\begin{equation}
 \rho_{j\geq2} (1) = \rho_j(0)\frac{v_j^{(2)}(0)- v_1^{(2)}(0)}{2- v_1^{(2)}(0)}.
\end{equation}
Knowing that the speed $v_j^{(l)}(0)$ in the sector 0 (equivalent to the homogeneous state)
is independent of $j$ for $j\geq l$, we have $v_{j\geq2}^{(2)}(0)=v_2^{(2)}(0)$ and we can write
\begin{equation}
 \rho_{j\geq2} (1) = \rho_j(0)\frac{v_2^{(2)}(0)- v_1^{(2)}(0)}{2- v_1^{(2)}(0)}.
\end{equation}
Together with $\rho_1(1)=0$ the above equation gives the soliton content of the sector 1 in terms of known properties of the homogeneous state. Finally we get the ball density in that sector:
\begin{equation}
h(1)=\sum_{j\ge1}j\rho_j(1)={v_2^{(2)}(0)-v_1^{(2)}(0)\over 2-v_1^{(2)}(0)  } \sum_{j\ge2}j\rho_j(0).
\end{equation}
An explicit calculation of the above sum gives finally
\begin{equation}
h(1)={z^4+z^3+2z^2\over z^4+z^3+4z^2+z+1 },
\label{eq:h1}
\end{equation}
in agreement with the numerical results of the left panel of figure~\ref{fig:l=2}. The heights and positions of the plateaux for $p_L>0$ and $p_R=0$ and an arbitrary value of $l$ will be given in the next subsection (Sec.~\ref{ssec:pL_only}).
The reversed case ($p_L=0$ and $p_R>0$) is discussed in Sec.~\ref{ssec:pR_only}. 

%--------------------------------------------------
\subsection{Explicit plateaux solutions for $0< p_L <  \frac{1}{2}$ and $p_R=0$}\label{ssec:pL_only}
%--------------------------------------------------

We again use the parameter $z$ related to $p_L$ by $p_L = \frac{z}{1+z}\, (0 < z < 1)$ and 
the shorthand
\begin{equation}
[j] = 1-  z^j
\end{equation}
when the formula is bulky.
Let us number the plateaux as $k= 0,1,2,\ldots$ from the left to the right, where the leftmost 
$0$-th one is of height $p_L$.
We employ the GHD equations in the convention of Remark \ref{re:M}.
Thus we use the velocity of the $j$-soliton $v_j(k)$, the total density 
$r_j(k)=\rho_j(k) + \sigma_j(k)$ and the occupancy 
$n_j(k) = \frac{\rho_j(k)}{\rho_j(k) + \sigma_j(k)}$
which satisfy $\rho_j(k) = n_j(k) r_j(k)$ for the $k$-th plateau.
The $(n_j(k))_{j \ge 1}$ for the $k$-th plateau is given by
\begin{align}\label{nL}
n_j(k)  = \theta(j>k) \frac{z^j(1-z)^2}{(1-z^{j+1})^2}.
\end{align}
See (\ref{teta}) for the definition of $\theta$.
For the time evolution $T_\infty$, the GHD equations read as 
\begin{align}
r_i(k)  &= 1 - \sum_{j=1}^\infty \tilde{M}_{ij}n_j(k) r_j(k) ,
\label{rL}
\\
r_i(k) v_i(k) & = i - \sum_{j=1}^\infty \tilde{M}_{ij}n_j(k) r_j(k) v_j(k),
\label{vL}
\end{align}
where $\tilde{M}_{ij} = (M-1)_{ij} = 2\min(i,j) - \delta_{ij}$
according to (\ref{Mdef}) and Remark \ref{re:M}.

For the time evolution $T_l$, (\ref{rL}) remains the same, whereas the first term $i$ on the RHS of (\ref{vL})
is replaced by $\min(l,i)$ (see also (\ref{vitesse0})).
The solution of (\ref{rL}) is given by 
\begin{align}
r_j(k)  &= \frac{[2k+3]+(2k+1-2j)[1]z^{k+1}}{[2k+3]+(2k+1)[1]z^{k+1}}
\qquad \quad(1 \le j  \le k),
\\
&= \frac{[k+1][k+2][2j+2]}{[j][j+2]([2k+3]+(2k+1)[1]z^{k+1})}
\qquad (j>k).
\end{align}
The solution to (\ref{vL}) is given by 
\begin{align}
v_j(k)  &= \frac{j [k+1][k+2]}{
\left(1+z^{k+1}\right) [k+2] + 2(k-j)[1]z^{k+1}}\quad (1 \le j\le k),
\label{eq:vjkA}
\\
&= \frac{\Gamma_j(k)}{(1+z^{j+1})[k+1][k+2]}\qquad\qquad\qquad \;\;(j>k),
\label{eq:vjkB}
\end{align}
where $\Gamma_j(k)$ is defined by 
\begin{equation}\begin{split}
\Gamma_j(k) &= 2k^2z^{k+2}[j]-2(k+1)^2z^{k+1}[j+2]-2z^{2k+3}[j-2k-2]\\
&-\frac{4z^{k+2}[j-k][k+1]}{[1]}+j(1+z^{j+1})\left([2k+3]+(1+2k)[1]z^{k+1}\right).
\end{split}
\end{equation}

From these results, the height $h(k)$ of the $k$-th plateau is calculated as
\begin{align}\label{hkL}
h(k) &= \sum_{j=1}^\infty j \rho_j(k) =  \sum_{j=k+1}^\infty j n_j(k) r_j(k)
= \frac{z^{k+1}([k+2]+k[1])}{[2k+3]+(2k+1)[1]z^{k+1}}.
\end{align}
This certainly satisfies $h(0) = \frac{z}{1+z}=p_L$.
And evaluating the above height for $k=1$ gives back (\ref{eq:h1}).

The position $\zeta(k)$ of the boundary of the $(k-1)$-th and the $k$-th plateaux is 
\begin{equation}
\zeta(k) = v_k(k) = \frac{k(1-z^{k+1})}{1+z^{k+1}} 
\label{eq:vk_inf}
\end{equation}
for $k \ge 1$.
One can check from (\ref{eq:vjkA}) and (\ref{eq:vjkB}) that $v_k(k-1) = v_k(k)=\zeta(k)	$  for $k\geq 1$,
in agreement with the discussion given in Sec.~\ref{ssec:dwghd}.

For $T_l$ with finite $l$, the height of the $k$-th plateau is given by  $\theta(k<l)h(k)$ 
with $h(k)$ still given by (\ref{hkL}).
On the other hand, the position $\zeta(k)$ gets modified into
\begin{align}\label{eq:zkl}
\zeta^{(l)}(k) = \frac{k(1-z^{k+1})(1+z^{l+1})}{(1+z^{k+1})(1-z^{l+1})}\qquad (1 \le k \le  l).
\end{align}

\begin{figure}\begin{center}
\includegraphics[width=\textwidth]{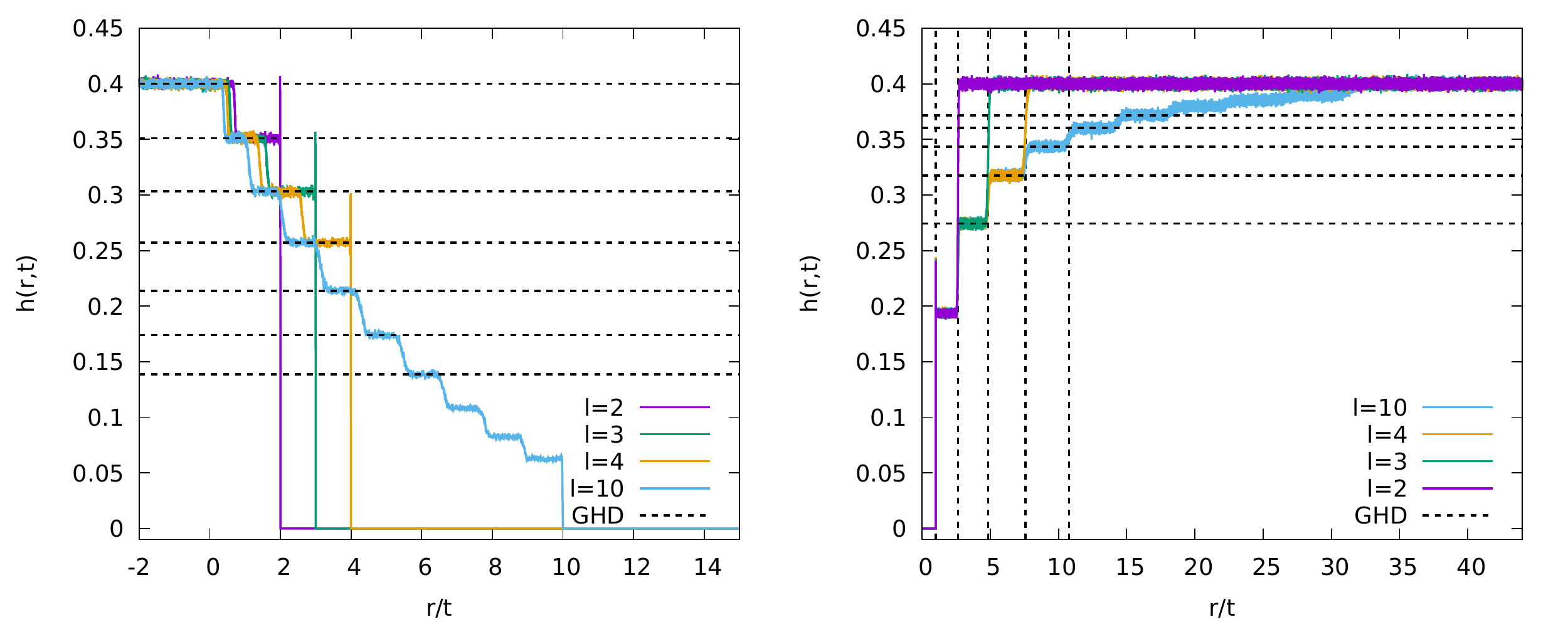}
\caption{Left: Ball density plotted as a function $r/t$ for the carrier capacities $l=2,3,4$  and 10, at fixed $t=500$. Initial ball densities: $p_L=0.4$ and $p_R=0$. The ball density is zero for $r/t>l$.
One can see that the density of the first plateau ($h(1)\simeq 0.35$) is the same for $l=2,3,4$ and 10. Similarly, the second plateau has the same density for $l=3,4$ and 10, etc. This fact is well reproduced in the GHD approach (Sec.~\ref{ssec:pL_only}).
Right: Same with $p_L=0$ and $p_R=0.4$ (for $t=1000$). Now we observe that not only the plateau heights, 
but also the plateau positions (velocities) are independent of $l$ (vertical dotted lines), in agreement with the results
of Sec.~\ref{ssec:pR_only}.
Note that the last plateau for $l=10$ starts at $\zeta(10)\simeq 31.2866$, a velocity which is significantly larger than the largest {\it bare} velocity $l=10$.
} \label{Fig:l=2_3_10_p=0.4}
\end{center}\end{figure}

\begin{figure}
\includegraphics[width=\textwidth]{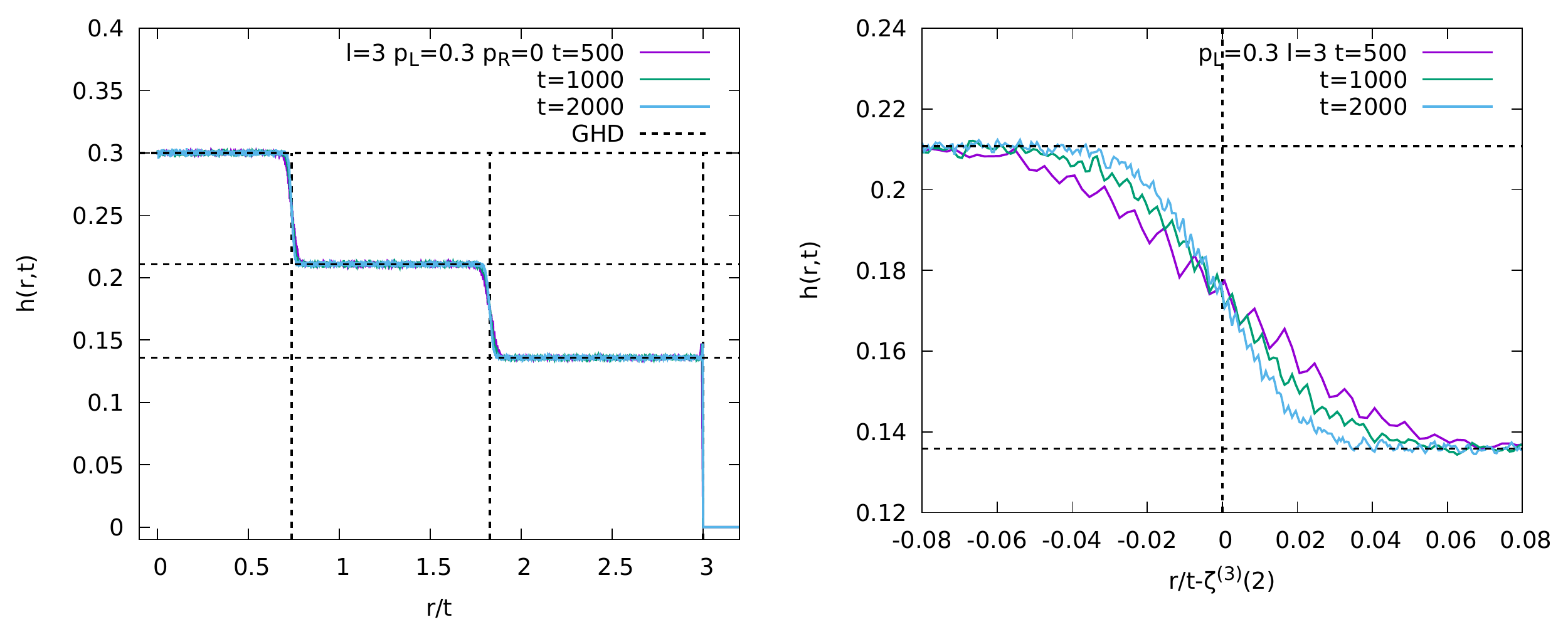}\begin{center}
\caption{Left: Ball density plotted as a function $\zeta=r/t$ for $l=3$ and different times.
Initial ball densities: $p_L=0.3$ and $p_R=0$.
The horizontal dotted line represents the GHD prediction for the ball densities in each plateau (\ref{hkL}) and the vertical dotted lines
represent  the GHD prediction for the end points of the plateaux (\ref{eq:vk_inf}).
Here $N_{\rm samples}=2.10^5$ and $L=2.10^5$.
Right: Zoom on the vicinity of the transition from the first to the second plateau at $\zeta^{(3)}(2)\simeq1.827$ (\ref{eq:zkl}).
The erratic variations reflect the statistical fluctuations ($N_{\rm samples}$ is large but finite).
The quantitative analysis of the (diffusive) broadening of the transition is carried out in Sec.~\ref{ssec:transition_widths}
} \label{Fig:l=3}
\end{center}\end{figure}

\begin{figure}\begin{center}
\includegraphics[width=0.6\textwidth]{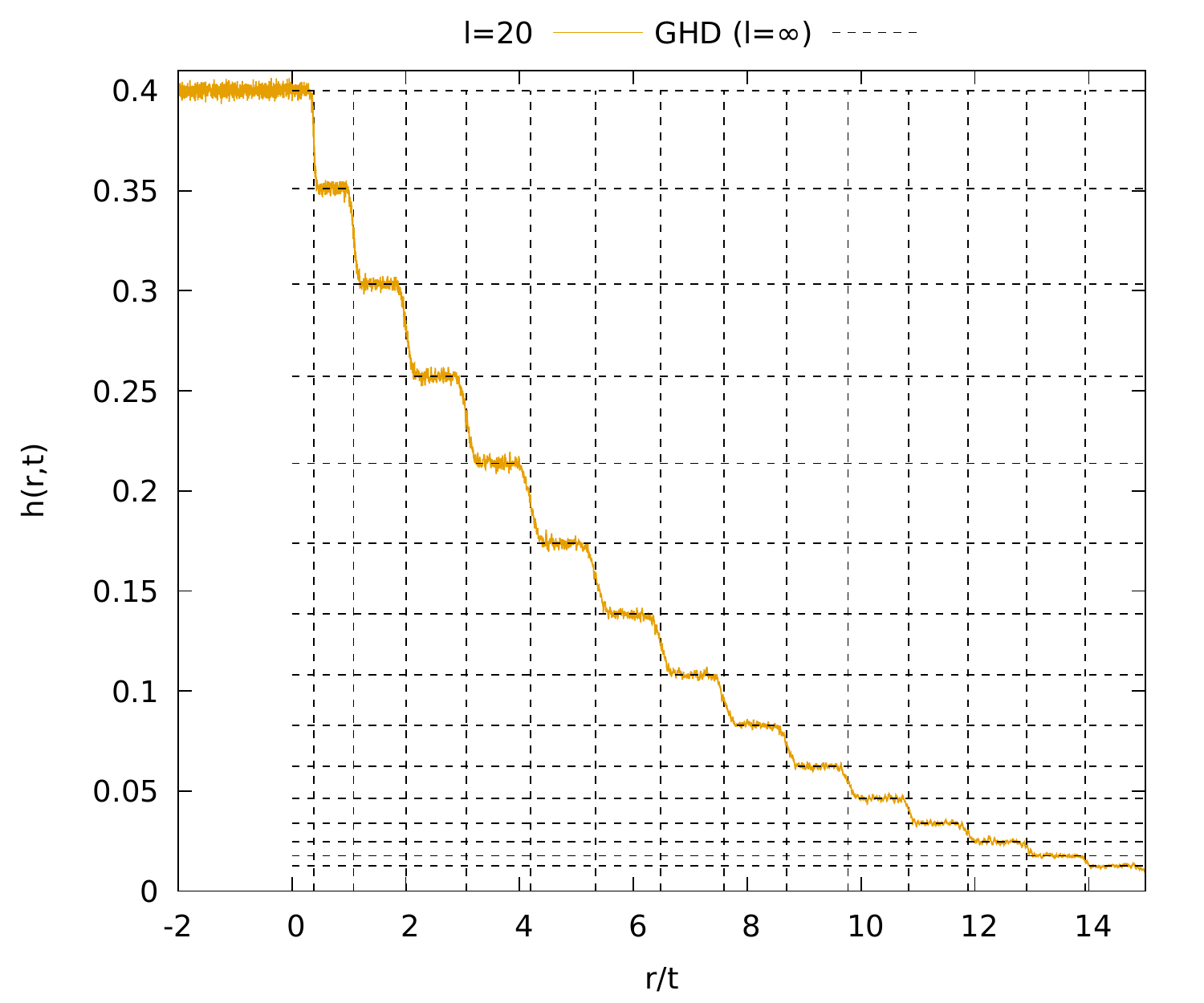}
\caption{Ball density for the capacity $l=20$ at fixed $t=500$. Initial ball densities: $p_L=0.4$  and $p_R=0$. The dotted lines correspond to the horizontal and vertical positions of the plateaux predicted by the GHD approach in the limit $l=\infty$,
see (\ref{hkL}) and (\ref{eq:vk_inf}).
} \label{fig:l=20pL}
\end{center}\end{figure}

The figures \ref{Fig:l=10_pR=0}, \ref{FigSolitonDensities}, the left panel of figure~\ref{fig:l=2}, the left panel of figure~\ref{Fig:l=2_3_10_p=0.4},  figure~\ref{Fig:l=3} and figure~\ref{fig:l=20pL} correspond to situations  $p_L>0$ and $p_R=0$, as discussed above. The heights of the plateaux turn out to be independent of $l$, but their positions depend on $l$. These heights and positions are in perfect agreement with the GHD results of eqs.~(\ref{hkL}) and (\ref{eq:zkl}).

%--------------------------------------------------
\subsection{Explicit plateaux solutions for $p_L=0$ and $0< p_R <  \frac{1}{2}$}\label{ssec:pR_only}
%--------------------------------------------------

Set $p_R = \frac{z}{1+z}$.
Let us number the plateaux as $k=0,1,2,\ldots$ from the left to the right, where the leftmost 
$0$-th one is of height $0$.
Then the occupation function of the $k$-th plateau is 
\begin{align}\label{nR}
n_j(k) = \theta(j\le k) \frac{z^j(1-z)^2}{(1-z^{j+1})^2},
\end{align}
which is complementary to (\ref{nL}).
The GHD equations are formally the same as (\ref{rL})  and (\ref{vL}) 
except that  $n_j(k)$ is now specified as  (\ref{nR}) instead of (\ref{nL}).
Note that the sums in (\ref{rL}) and (\ref{vL}) over $j$ are reduced to the finite one over $1\le j \le k$.
The solution to (\ref{rL}) on $r_j(k)$ is
\begin{align}
r_j(k)  &= \frac{[1][j+1]X_j(k)}
{(1+z)[j][j+2]([2k+3]-(2k+3)[1]z^{k+1})}\quad (1 \le j \le k),
\label{cdknki}\\
&= \frac{[1][k+1][k+2]}{(1+z)([2k+3]-(2k+3)[1]z^{k+1})}
\quad (j>k).
\end{align}
The solution to (\ref{vL})  on the velocity $v_j(k)$ is
\begin{align}
v_j(k) &= \frac{(1+z)[k+1][k+2]\bigl(j[j+2]-(j+2)[j]z\bigr)}
{[1]^2X_j(k)}
\quad (1 \le j \le k),
\label{eq:vjkC}
\\
&= \frac{(1+z)\bigl((2k(k+2)-j(2k+3))[1]^2z^{k+1}-2z[k][k+2]
+j[1][2k+3]\bigr)}
{[1]^2[k+1][k+2]}
\quad (j > k),
\label{eq:vjkD}
\end{align}
where $X_j(k)$ in (\ref{cdknki}) and (\ref{eq:vjkC}) is defined by
\begin{align}
X_j(k) = (1+z^{j+1})[2k+3]+(2k-2j-1)[j]z^{k+2}-(2k-2j+3)[j+2]z^{k+1}.
\end{align}
From these results, the height $h(k)$ of the $k$-th plateau is calculated as
\begin{align}\label{hkR}
h(k) &= \sum_{j=1}^\infty j \rho_j(k) =  \sum_{j=1}^kj n_j(k) r_j(k)
= \frac{z([2k+2]-(k+1)[2]z^k)}{(1+z)\bigl([2k+3]-(2k+3)[1]z^{k+1}\bigr)}.
\end{align}
This certainly satisfies $h(0)=0$ and $h(\infty) = \frac{z}{1+z} = p_R$.

The position $\zeta(k)$ of the boundary of the $(k-1)$-th and the $k$-th plateaux is 
\begin{align}\label{zetaR}
\zeta(k) = v_k{(k)} =  
\frac{1+z^{k+1}}{1-z^{k+1}}
\left(k\frac{1+z}{1-z} -\frac{2z(1+z)(1-z^k)}{(1-z)^2(1+z^{k+1})}\right) = 
\left. v^{(k)}_k \right|_{a=z}\text{ in (\ref{mirei})}.
\end{align}
In particular one has $\zeta(1)=1$.
And here also one can check from (\ref{eq:vjkC}) and (\ref{eq:vjkD}) that $v_k(k-1) = v_k(k)=\zeta(k)$  for $k\geq 1$.

It is also instructive to consider (\ref{hkR}) and (\ref{zetaR}) in the half-filled limit $z\to1$. There one finds
\begin{equation}
 \lim_{z\to1} h(k)=\frac{k}{2k+3}
\end{equation}
and
\begin{equation}
 \lim_{z\to1} \zeta(k)=\frac{1}{3}k(k+2),
\end{equation}
with a step position $\zeta(k)$ which turns out to grow much faster than the bare velocity $k$.

For $T_l$ with finite $l$, the results
(\ref{hkR})  remains valid for $0 \le k \le l-1$ and  $h(l)= \frac{z}{1+z}=p_R$.
The result  (\ref{zetaR}) is valid for $1 \le k \le l-1$.

The right panel of figure~\ref{fig:l=2} and the right panel of figure~\ref{Fig:l=2_3_10_p=0.4} describe such situations, where $p_L=0$ and $p_R>0$. In such cases, the heights {\em and} the positions (or velocities) of the plateaux indeed appear to be independent of $l$, as expected from the GHD results presented above.

%--------------------------------------------------
\subsection{Transitions between plateaux and diffusive broadening of the steps}
\label{ssec:transition_widths}
%--------------------------------------------------

\begin{figure}
\includegraphics[width=\textwidth]{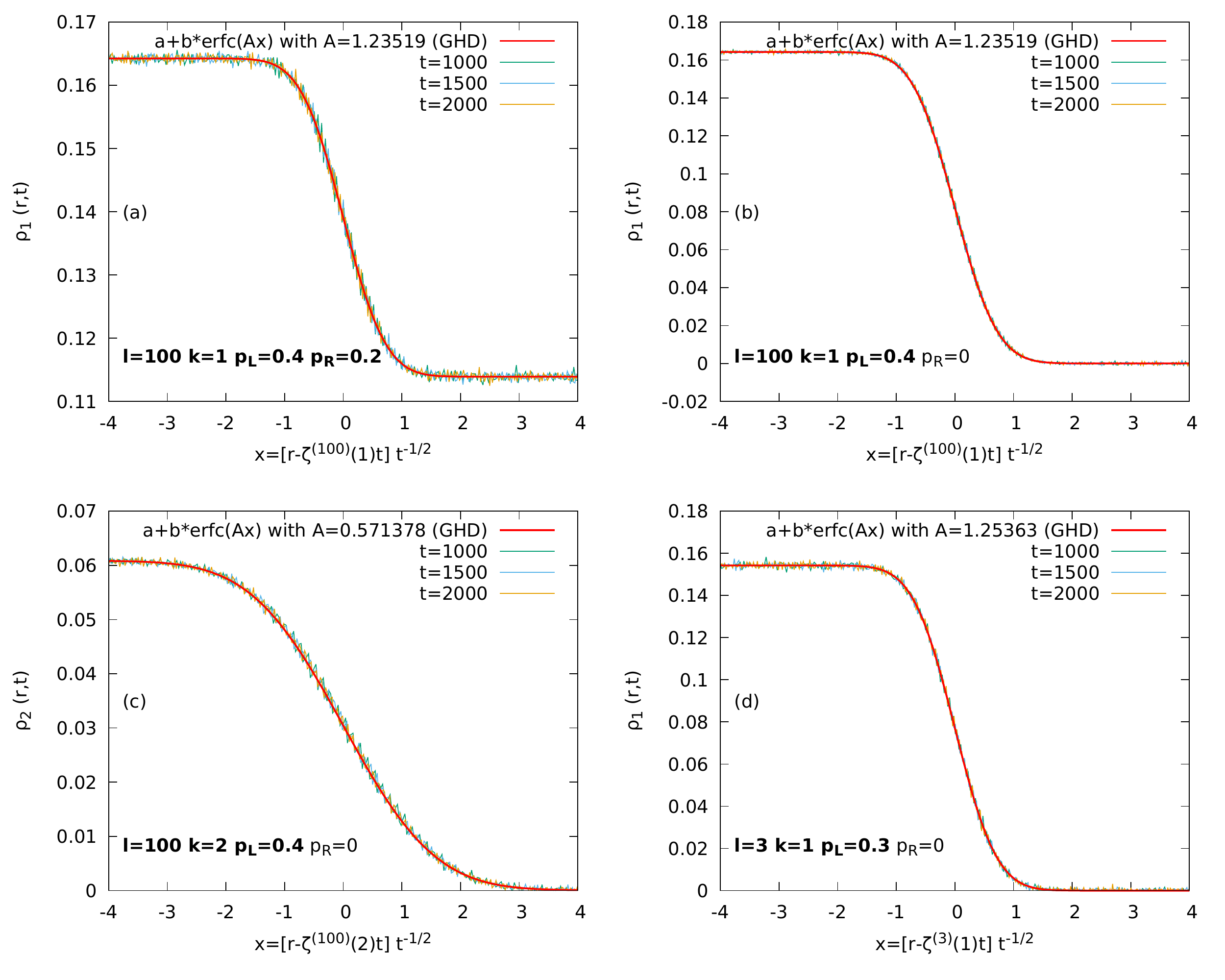}\begin{center}
\caption{Zoom on the soliton density $\rho_j(r,t)$ at four different steps between plateaux.
The red curves are the prediction of the diffusive corrections to GHD (\ref{fluct3}).
In all cases the horizontal axis corresponds to $\left[ r-\zeta^{(l)}(k)t\right]t^{-1/2}$ and provides a good
collapse of the curves measured at different times ($t=500$, 1500 and 2000). This shows that
the scaling of the transition regions is diffusive.
The high precision was achieved thanks to simulations done up to $t=2000$ with $L=10^6$ sites and $N_{\rm samples}=10^6$, corresponding to $\sim 10^{15}$ particle moves.
The four panels correspond to different parameters of the model:
(a) $p_L=0.4$, $p_R=0.2$, $l=100$ and $k=1$.
(b) $p_L=0.4$, $p_R=0$, $l=100$ and $k=1$.
(c) $p_L=0.4$, $p_R=0$, $l=100$ and $k=2$.
(a) $p_L=0.3$, $p_R=0$, $l=3$ and $k=1$.
The inverse width $A=1/\sqrt{2\Sigma^2}$ in panel (a) was obtained by computing numerically $\Sigma_k^2$ in (\ref{fluct3}).
In panels (b), (c) and (d), $A^2$ is calculated from (\ref{Sig}) with 
$(k,l) = (1,100), (2,100)$ and $(1,3)$.
The simulations and the theoretical values of $\Sigma_k^{(l)}$ agree with a $10^{-2} \sim 10^{-3}$ relative accuracy, which is the order of magnitude of the statistical fluctuations on the simulation side. 
} \label{Fig:width}
\end{center}\end{figure}

At this stage the GHD approach predicts sharp ({\it i.e.} discontinuous) steps/transitions between plateaux in the variable $\zeta$.
However, in the numerical simulations, the steps in the density curves (ball density or soliton densities) exhibit some visible widths. This can be seen, for instance, in the right panel of figure~\ref{Fig:l=3}.
A careful analysis of the time-dependence of these transition regions, as proposed in figure~\ref{Fig:width}, shows that
the spatial extent of these regions is $\sim \sqrt{t}$ in the variable $r$, corresponding to a diffusive behavior.

The finite widths of the plateau transitions can be interpreted as the fact that some
solitons have traveled faster or more slowly than the mean velocity predicted by GHD. This is to be expected, as the soliton densities are fluctuating from one initial configuration to another, and these density fluctuations induce velocity fluctuations.
If one assumes that the density fluctuations that a given tagged soliton 
``sees'' are uncorrelated, its mean velocity computed over a certain time will
be distributed in a Gaussian way (for long enough time). This should lead to a diffusive broadening of the transitions between 
consecutive plateaux, as observed numerically.
We stress that the dynamics is here completely deterministic, and the diffusion originates from the randomness in the initial conditions. 

In the following we explain how to describe quantitatively this diffusive broadening.
We will describe how to compute the  shape of the density curves joining two consecutive plateaux.
There is already a abundant literature on diffusive corrections to GHD~\cite{doyon_dynamics_2017,GHKV,DBD,GV19,NBD19}
and the argument we propose below in the context of the BBS is close to the one given in \cite{GHKV}.

Across the characteristic (wave front)
between plateau $k-1$ and $k$, the pseudoenergy $\epsilon_k$  defined by $y_k=\exp{(-\epsilon_k)}$ changes abruptly from  $\epsilon_k(k-1)=(\epsilon_L)_k$ to $\epsilon_k(k)=(\epsilon_R)_{k}$. 
At position $r$ and time 
$t$, it takes the value $(\epsilon_L)_k$ or $(\epsilon_R)_k$ according to whether it is located left or right of the fluctuating wave front. In other words, its value depends if the average velocity of the wave-front position, $\bar v_k=(1/t)\int_{0}^t v_k(s) ds$, is larger or smaller than $r/t$.

The above wave-front velocity fluctuates due to fluctuations in the background of solitons $ i\ne k$ crossing it.
For a given random initial condition these background fluctuations can be described by pseudoenergy density fluctuations $\delta\bar\epsilon_i$.
The fluctuations of $\bar v_k$, denoted by $\delta \bar v_k$, can thus be written as
\begin{equation}
\delta \bar v_k=\sum_{i\ne k}\left({\partial v_k\over \partial \epsilon_i}\right)\delta\bar\epsilon_i.
\label{fluct1}
\end{equation}
In the above expression, the pseudoenergy density fluctuation $\delta\bar\epsilon_i$ is
obtained by averaging the pseudoenergy over the length $t |v_i-v_k|$, which is
the distance the wave front has traveled in a (moving) frame where the background of $i$-solitons is at rest.
Assuming that the pseudoenergy fluctuations at different point in space are uncorrelated, we find that at long times $\delta\bar v_k$ follows a Gaussian distribution with a variance given by:
\begin{equation}
\langle (\delta \bar v_k )^2\rangle = \sum_{i\ne k}\left({\partial v_k\over \partial \epsilon_i}\right)^2{\langle (\delta  \epsilon_i)^2\rangle \over |v_i-v_k|t},
\label{fluct2}
\end{equation}
where $\langle (\delta  \epsilon_i)^2\rangle $ is the stationary (and homogeneous) $i$-soliton pseudoenergy variance per site. 

This quantity $\langle (\delta  \epsilon_i)^2\rangle$   can be obtained by a thermodynamic calculation, by expanding 
the free energy per site (\ref{fp}) at quadratic order in $\epsilon$ \cite{FS}.
From (\ref{Fdef}) and (\ref{tba}), the free energy derivative is given by:
\begin{equation}
\frac{\partial {\mathcal F}}{\partial \epsilon_i}
= \sum_{n=1}^\infty X_n  \frac{\partial \rho_n}{\partial \epsilon_i},
\label{F=Xro}
\end{equation}
where $X_n=\frac{\partial {\mathcal F}}{\partial \rho_n}$ is given by the TBA equation (\ref{tba})
\begin{equation}
X_n = \sum_{p=1}^\infty\min(n,p)\beta_p
-\log(1+\mathrm{e}^{\epsilon_n}) + 2\sum_{p=1}^\infty\min(n,p) \log(1+\mathrm{e}^{-\epsilon_p})
\label{Xdef}
\end{equation}
and the  free energy Hessian is given by:
\begin{equation}
\frac{\partial^2 {\mathcal F}}{\partial \epsilon_k\partial \epsilon_p}
= \sum_n \frac{\partial X_n}{\partial \epsilon_p}
\frac{\partial \rho_n}{\partial \epsilon_k}, 
\label{hess}
\end{equation}
where we have taken into account that it is evaluated at the minimum of the free energy, where $X_k=0$.
Using the pseudoenergies $\sigma=\rho/y=\mathrm{e}^{\epsilon}\rho$, (\ref{eq:def_sigma}) can be rewritten as:
\begin{align}
1=(M+\mathrm{e}^{\epsilon})\rho.
\end{align}
Differentiating with respect to $\rho_n$ we deduce:
\begin{equation}
{\partial \epsilon_j\over \partial \rho_n}=-\sigma_j^{-1}(M+\mathrm{e}^{\epsilon})_{jn}
\label{Xro1}
\end{equation}
and from (\ref{Xdef}):
\begin{equation}
{\partial X_n\over \partial \epsilon_p}=-{1\over 1+\mathrm{e}^{\epsilon_p}}(M+\mathrm{e}^{\epsilon})_{pn}.
\label{Xro2}
\end{equation}
Substituting the inverse of (\ref{Xro1}) and (\ref{Xro2}) in (\ref{hess}) we obtain: 
\begin{align}
\frac{\partial^2 {\mathcal F}}{\partial \epsilon_k\partial \epsilon_p}
&=\delta_{kp} \frac{\sigma_k}{1+ \mathrm{e}^{\epsilon_k}}.
\end{align}
This yields diagonal pseudoenergy fluctuations:\footnote{This is a spectral parameter-free version of 
\cite[eq.(A.1)]{FS}.
A similar result was originally given in \cite{YY69} for the density correlations of the single-component boson.}
\begin{align}\label{fs}
\langle \delta \epsilon_k \delta \epsilon_p \rangle = \delta_{kp}
\frac{1+\mathrm{e}^{\epsilon_k}}{\sigma_k}.
\end{align}
Now, coming back to (\ref{fluct2}), we need to compute the $\partial v_k /\partial \epsilon_i$ terms.
These velocity derivatives can be obtained from (\ref{vitess2}), $v=\kappa^{\rm dr}/1^{\rm dr}$,
which can be written as $v=\left((1+M{\bf y})^{-1}\kappa\right)/\left((1+M{\bf y})^{-1}1\right)$.
The result is:
\begin{equation}
{\partial v_k\over \partial \epsilon_i}= \beta_{ki} {\rho_i\over \sigma_k} (v_k-v_i),
\label{vites'}
\end{equation}
where $\beta$ is the matrix $\beta=(1+M{\bf y})^{-1}M$.
From the definition (\ref{dressed2}) of the dressing operation, $\beta$ corresponds to $M^{\rm dr}$.

Putting (\ref{fluct2}), (\ref{fs}) and (\ref{vites'}), together, we obtain:%
\footnote{With (\ref{diagmod}) we obtain an equivalent expression  $\Sigma_k^2=\sum_{i}{\tilde \beta_{ki}^2 \over r_k^2}|v_k-v_i| r_i n_i(1-n_i)$ with $\tilde \beta=(1+\tilde M{\bf n})^{-1}\tilde M$.}
\begin{equation}
t\langle (\delta \bar v_k )^2\rangle =  \Sigma_k^2=\sum_{i}{\beta_{ki}^2 \over \sigma_k^2}|v_k-v_i| \sigma_i y_i(1+y_i).
\label{fluct3}
\end{equation}

Since we expect a Gaussian distribution of the front-velocity fluctuations in the long time limit, the above variance
is essentially enough to characterize the distribution of the front position. 
Let us consider a local quantity, the density of $j$-solitons, evaluated at
$(r,t)$ for a value $\zeta=r/t$ close to the mean position $\zeta(k)$ of the step number $k$.
For realizations such that the front has a velocity $\bar v_k$ larger than $\zeta$, the $j$-soliton density will be equal to its value $\rho_j(k-1)$
in the plateau $k-1$. On the other hand, for realizations such that the front has a velocity $\bar v_k$ smaller than $\zeta$,  the $j$-soliton density will be equal to its value $\rho_j(k)$
in the plateau $k$. Since $\bar v_k$ is distributed in a Gaussian way,
the probability to be above (or below) a certain value $r/t$ can be written simply with the complementary error function
$\text{erfc}(u) = \frac{2}{\sqrt{\pi}}\int_u^\infty \mathrm{e}^{-s^2}ds$.
The Gaussian being characterized by the mean $\zeta(k)$ and 
the variance $\Sigma_k^2$ (\ref{fluct3}), the realization-averaged soliton densities in the vicinity of the step $k$ reads:
\begin{equation}
\langle\rho_j(r,t)\rangle=\frac{1}{2}\left(\rho_j(k-1)-\rho_j(k)\right) {\rm erfc}\left(\sqrt{t\over 2} {r/t-\zeta(k) \over \Sigma_k}\right)+\rho_j(k).
\label{shape}
\end{equation}
And the above form should in fact hold for any local quantity which is a function of the pseudoenergies.

\begin{figure}\begin{center}
\includegraphics[width=\textwidth]{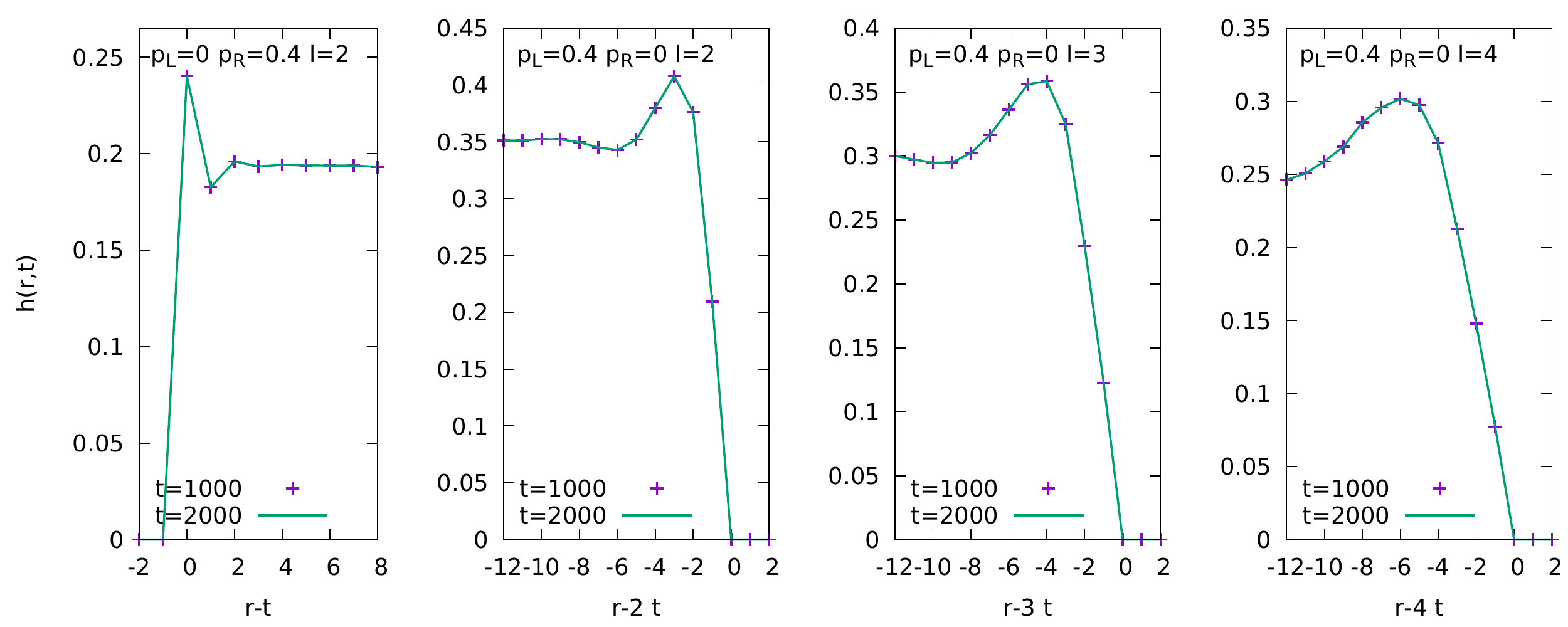}
\caption{
Ball density in the vicinity of a step to (or from) a zero-density plateau when $p_L$ or $p_R$ is set to zero. Contrary to the others, these 
steps do not broaden diffusively since there is only one species of soliton on the nontrivial side of the step. Instead, these steps show a peak (or spike) whose spatial extent is constant in time. From left to right:
i) $l=2$  $p_L=0$ $p_R=0.4$ $\zeta=1$
ii) $l=2$  $p_L=0.4$ $p_R=0$ $\zeta=2$
iii) $l=3$  $p_L=0.4$ $p_R=0$  $\zeta=3$
iv) $l=4$  $p_L=0.4$ $p_R=0$  $\zeta=4$.
Note that the curve at different times are superimposed.
} \label{FigPeak}
\end{center}\end{figure}

%--------------------------------------------------

We finish this subsection by giving a conjectural formula for the diffusive step width $\Sigma_k$ 
for the plateaux generated from the  initial condition 
$p_L>0$ and $p_R=0$ by the time evolution $T_l$.
We denote it by $\Sigma^{(l)}_k$ exhibiting the dependence on $l$.
It was obtained by computing
successive approximations as a function of the initial state fugacity $z=z_L$,
using a computer algebra software and truncations of the matrix $M$ at increasing orders. 
\begin{align}\label{Sig}
2\bigl(\Sigma^{(l)}_k\bigr)^2 = 
\frac{8k^2z^{k+1}(1-z^{k+1})(1-z^{l-k})(1+z^{l+k+2})}{(1+z^{k+1})^3(1-z^{l+1})^2}
\qquad (1 \le k \le l).
\end{align}
Note that for finite $l$, there are only $l$ plateaux with non-zero height.
The leftmost step corresponds to $\Sigma^{(l)}_1$, and 
the rightmost one $\Sigma^{(l)}_l$ vanishes reflecting the no diffusive broadening 
mentioned in the caption of figure \ref{FigPeak}.
We checked that (\ref{Sig}) matches the numerical solutions of the GHD equations at a few fixed $p_L$.
As shown in figure~\ref{Fig:width}, they are also in very good agreement with the simulations.

\newpage

%--------------------------------------------------
\section{Outlook}\label{sec:discussion}
%--------------------------------------------------

In this paper we have exclusively treated the most basic BBS with only one kind of balls. 
It is natural to extend the analysis to the 
BBS with $n$ kinds of balls \cite{Ta93} which is known to be 
associated with the quantum group $U_q(\widehat{sl}_{n+1})$.
Here is an example of three soliton scattering with amplitude 5, 3 and 2 for $n=3$:

\vspace{0.1cm}
\begin{center}
$ \ldots 0003222100002110031000000000000000000000000000000 \ldots $ 
$ \ldots 0000000032221002110310000000000000000000000000000 \ldots $ 
$ \ldots 0000000000000321002203211100000000000000000000000 \ldots $ 
$ \ldots 0000000000000000321022000032111000000000000000000 \ldots $ 
$ \ldots 0000000000000000000310222000000321110000000000000 \ldots $ 
$ \ldots 0000000000000000000003100222000000003211100000000 \ldots $ 
$ \ldots 0000000000000000000000031000222000000000032111000 \ldots $
\end{center}

\vspace{0.1cm}
As one can observe, the amplitude of solitons are again individually conserved 
before and after the collisions. 
A new aspects here is that they now possess the internal degrees of freedom 
which are nontrivially exchanged like quarks in hadrons. 
It is an interesting outstanding problem to seek   
a speed equation for such a system with non-diagonal scattering,
and more broadly, to formulate  a systematic higher rank (nested) extension of GHD
that fits the generalized BBS associated with quantum groups in general \cite{IKT12}.
We hope to report on these issues in a future work.

%\section*{Appendix}
\appendix

%--------------------------------------------------
\section{Transfer matrix formalism of GGE partition function of BBS}\label{app:tr}
%--------------------------------------------------

Introduce the matrices
\begin{align}
V^{(l)}_\eta &= \bigl(V^{(l)}_\eta(n,\tilde{n})\bigr)_{n,\tilde{n}=0}^l
\in \mathrm{Mat}(l+1,l+1) \quad (\eta \in \{0,1\}),
\\
V^{(l)}_\eta(n,\tilde{n}) & = 
\theta\bigl(n, \tilde{n} \,\text{and}\, \eta \;\text{fit } (\ref{vd})\bigr) 
\mathrm{e}^{-\beta_l \theta(\eta>\tilde{\eta})},
\label{vdef}
\end{align}
where $\theta$ is defined by (\ref{teta}).
In (\ref{vdef}),  $\tilde{\eta} \in \{0,1\}$ is also determined by the fitting 
diagram in (\ref{vd}).
The factor 
$\mathrm{e}^{-\beta_l \theta(\eta>\tilde{\eta})}$
 incorporates the  local Boltzmann weight  from the $l$-th energy $E_l$ (\ref{el}).
Now consider the partition function 
\begin{align}
{\tilde Z}_L(\beta_1,\ldots, \beta_r, \beta_\infty)
= \sum_{\eta \in \{0,1\}^L} \mathrm{e}^{-\beta_1E_1
-\cdots - \beta_r E_r - \beta_\infty E_\infty}.
\end{align}
This is equal to (\ref{zl})  with the corresponding choice $s$ of the temperatures 
except that the sum is not restricted by the condition mentioned under it.
In the rest of this appendix, we assume $L$ is odd\footnote{For even $L$, 
a correction term is necessary in (\ref{stoko}) to take into account of the fact that 
$c_l(\eta)$ satisfying (\ref{tldef}) is not unique at $\sum_{i\in \Z_L} \eta_i = L/2$
\cite[Prop.2.1]{KTT06}.}.
By the construction we have 

\begin{align}\label{stoko}
{\tilde Z}_L(\beta_1,\ldots, \beta_r, \beta_\infty)
&= \sum_{\eta_1,\ldots, \eta_L \in \{0,1\}}
\mathrm{Tr}(V^{(1)}_{\eta_1} \cdots V^{(1)}_{\eta_L}) \cdots 
\mathrm{Tr}(V^{(r)}_{\eta_1} \cdots V^{(r)}_{\eta_L})
\mathrm{e}^{-\beta_\infty(\eta_1+ \cdots + \eta_L)},
\end{align}
where $\mathrm{e}^{-\beta_\infty(\eta_1+ \cdots + \eta_L)}$ 
is the Boltzmann factor for $E_\infty$  in (\ref{ei}).
We have included it separately from $E_1,\ldots, E_r$ 
since it can be incorporated by a simple scalar exceptionally.
Define the transfer matrix $V$ by
\begin{align}\label{deema0}
V = V^{(1)}_0 \otimes \cdots \otimes V^{(r)}_0
+  \mathrm{e}^{-\beta_\infty}
V^{(1)}_1 \otimes \cdots \otimes V^{(r)}_1
\in \mathrm{Mat}((r+1)!, (r+1)!).
\end{align}
Then we have
\begin{align}\label{deema}
{\tilde Z}_L(\beta_1,\ldots, \beta_r, \beta_\infty)
 = \mathrm{Tr}(V^L).
\end{align}
This puts us in a standard situation, i.e., 
the free energy density in the thermodynamic limit 
is reduced to $-\log$ of the largest eigenvalue of the transfer matrix $V$.
We expect that in the limit $L \rightarrow \infty$, 
the above ${\tilde Z}_L(\beta_1,\ldots, \beta_r, \beta_\infty)$ yields the 
same free energy density as (\ref{zl}).

%\begin{example}
\par \vspace{0.3cm}\noindent{\bf Example A.1.} 
The transfer matrix for $\tilde{Z}_L(\beta_1, \beta_\infty)$ is 
\begin{align}
V = \begin{pmatrix} 1 & 0 \\ 1 & 0 \end{pmatrix}
+\mathrm{e}^{-\beta_\infty}\begin{pmatrix} 0 & \mathrm{e}^{-\beta_1} \\ 0 & 1 \end{pmatrix}.
\end{align}
By expressing $\beta_1, \beta_\infty$ in terms of $a,z$ according to (\ref{qlim}),
one finds that the largest eigenvalue of $V$ is $\frac{1-az}{1-a}$ 
in agreement with the free energy result (\ref{tatsuki}).
%\end{example}

%\begin{example}
\par \vspace{0.3cm}\noindent{\bf Example A.2.} 
The transfer matrices for $\tilde{Z}_L(\beta_1, \beta_2,\beta_\infty)$ is 
\begin{align}\label{vmat}
V = \begin{pmatrix} 1 & 0 \\ 1 & 0 \end{pmatrix} \otimes
\begin{pmatrix}1 & 0 & 0\\ 1 & 0 & 0 \\ 0 & 1 & 0 \end{pmatrix}
+ \mathrm{e}^{-\beta_\infty}
\begin{pmatrix} 0 & \mathrm{e}^{-\beta_1} \\ 0 & 1 \end{pmatrix} \otimes
\begin{pmatrix}0 & \mathrm{e}^{-\beta_2}  & 0\\ 0 & 0 &  \mathrm{e}^{-\beta_2} 
\\ 0 & 0 & 1 \end{pmatrix}.
\end{align}
The largest eigenvalue of this will be treated in Example B.2.
%\end{example}

%--------------------------------------------------
\section{Low temperature expansion for 
GGE$(\beta_1, \ldots, \beta_s)$}\label{app:A}
%--------------------------------------------------

Here we treat the general Y-system (\ref{y1})--(\ref{ys}).
It is written as
\begin{align}\label{ysys}
Y_i^2 &= \mathrm{e}^{\beta_i}\prod_{j=1}^s(1+Y_j)^{A_{ij}},
\end{align}
where the quantity $A_{ij}$ and the related $C_{ij}$ 
with its useful property are given by 
\begin{align}
&A_{ij}=2\delta_{ij}-C_{ij},
\quad C_{ij}= 2\theta(i=j<n)+\theta(i=j=n)-\theta(|i-j|=1),
\\
&\sum_{j=1}^s \min(i,j)C_{jk}=\delta_{ik},
\qquad 
\sum_{i=1}^sC_{ij}=\delta_{j,1},
\label{cinv}
\end{align}
where $\theta$ is defined in (\ref{teta}).
The matrix $(C_{ij})_{i,j=1}^s$ is the symmetrized Cartan matrix  
of $so(2s+1)$.
The Y-system (\ref{ysys}) is transformed into another 
difference equation called Q-system: 
\begin{align}\label{qm}
Y_i = \mathrm{e}^{\frac{\beta_i}{2}}\prod_{j=1}^s(\tilde{Q}_j)^{A_{ij}},
\quad
(\tilde{Q}_i)^2 = \mathrm{e}^{\frac{\beta_i}{2}}\prod_{j=1}^s(\tilde{Q}_j)^{A_{ij}}+1.
\end{align}
Further setting 
\begin{align}\label{wb}
\tilde{Q}_i = \mathrm{e}^{\frac{1}{2}\sum_{j=1}^s\min(i,j)\beta_j}Q_i,
\qquad
w_i = \mathrm{e}^{-\sum_{j=1}^s\min(i,j)\beta_j}
\end{align}
and using (\ref{cinv}),  the relations 
(\ref{qm}) are cast into
\begin{align}
Y_i = w_i^{-1}\prod_{j=1}^s(Q_j)^{A_{ij}},
\qquad
\prod_{j=1}^s(Q_j)^{-C_{ij}} + w_i Q^{-2}_i = 1.
\label{qgen}
\end{align}
Due to 
$1+Y^{-1}_i = \prod_{1\le j \le s} (Q_j)^{C_{ij}}$, 
the free energy (\ref{efe}) is expressed as
\begin{align}\label{fq1}
\mathcal{F} = -\sum_{i,j=1}^s C_{ij}\log Q_j = -\log Q_1.
\end{align}

The latter relation in (\ref{qgen}) 
exactly fits \cite[eq.(2.5)]{KNT02}
with $D_{ij}=-C_{ij}$ and $G_{ij}=-2\delta_{ij}$.
Therefore we have the power series formulas 
\cite[eq.(2.17)]{KNT02} and  
\cite[eq.(2.38)]{KNT02}.
In the present setting they read as 
\begin{align}
Q_1^{\mu_1}\cdots Q_s^{\mu_s}&= 
 \sum_{(m_1,\ldots, m_s)\in (\Z_{\ge 0})^s}
\!\!\!\!\!\!\!\! w_1^{m_1}\cdots w_s^{m_s}\Bigl(\,
\prod_{j\in H}\frac{(q_j+1)_{m_j-1}}{m_j!}\Bigr)
\det\bigl( F_{jk}\bigr)_{j,k \in H},
\label{qmu}\\
\log Q_1 &=  
\sum_{(m_1,\ldots, m_s)\in (\Z_{\ge 0})^s}
\!\!\!\!\!\!\!\! w_1^{m_1}\cdots w_s^{m_s}\Bigl(\,
\prod_{j\in H}\frac{(\bar{q}_j+1)_{m_j-1}}{m_j!}\Bigr)
\sum_{r \in H}\det\bigl( \bar{F}_{jk}\bigr)_{j,k \in H\setminus \{r\}},
\label{lgQ}
\\
q_j &= \sum_{k=1}^s\min(j,k)(\mu_k-2m_k),\quad
F_{jk} = \delta_{jk}q_j +2\min(j,k)m_k,
\\
\bar{q}_j&
= q_j|_{\mu_1=\cdots = \mu_s=0},
\qquad \qquad\quad\;\;
\bar{F}_{jk} = F_{jk}|_{\mu_1=\cdots = \mu_s=0},
\\
H &=\{1\le i \le s\mid m_i >0\},
\quad
(x)_n = x(x+1)\cdots (x+n-1)\quad (n \in \Z_{\ge 0}).
\end{align}
In (\ref{qmu}), the powers $\mu_1, \ldots, \mu_s \in \C$ 
are arbitrary\footnote{ $Q_i$'s are power series in $w_1,\ldots, w_s$ with 
a unit constant term for which their complex power is unambiguously defined.},
and (\ref{lgQ}) is deduced by differentiating (\ref{qmu}) with respect to 
$\mu_1$.  
These formulas are outcome of the theory of 
{\em generalized Q-systems}. 
It is the most systematic synthesis of numerous preceding results on the constant TBA 
equations for XXX, XXZ type spin chains 
and the Sutherland-Wu equations for ideal gas with Haldane statistics.
See for example \cite[Sec.~2.4]{KNT02} for a historical account.

From (\ref{zl})--(\ref{scal}), (\ref{wb}) and (\ref{fq1})  
one can derive the energy density as
\begin{align}\label{elogq}
\varepsilon_i &= \frac{\partial {\mathcal F}}{\partial \beta_i}
= -\sum_{j=1}^s \min(i,j) w_j\frac{\partial {\mathcal F}}{\partial w_j}
= \sum_{j=1}^s\min(i,j) w_j \frac{\partial \log Q_1}{\partial w_j}.
\end{align}
Comparing (\ref{elogq}) with the left relation in (\ref{epsi})
we get the density of $i$-solitons:
\begin{align}\label{toko}
\rho_i = w_i \frac{\partial \log Q_1}{\partial w_i}.
\end{align}
Substituting (\ref{lgQ}) into (\ref{toko}) 
we obtain 
\begin{align}\label{cdksesui}
\rho_i &=
\sum_{(m_1,\ldots, m_s)\in (\Z_{\ge 0})^s}
\!\!\!\!\!\!\!\! m_i\, w_1^{m_1}\cdots w_s^{m_s}
\Bigl(\,
\prod_{j\in H}\frac{(\bar{q}_j+1)_{m_j-1}}{m_j!}\Bigr)
\sum_{r \in H}\det\bigl( \bar{F}_{jk}\bigr)_{j,k \in H\setminus \{r\}}.
\end{align}
In view of  
$w_1^{m_1}\cdots w_s^{m_s} 
= \mathrm{e}^{-\sum_{1\le j,k \le s}\min(j,k) \beta_j m_k}$,
the series (\ref{qmu}), (\ref{lgQ}) and 
(\ref{cdksesui}) are low-temperature expansions. 
They have a finite convergence radius mentioned after 
\cite[eq.(2.15)]{KNT02}.
To investigate the behavior around them 
is beyond the scope of this paper.
 
%\begin{example}
\par \vspace{0.3cm}\noindent{\bf Example B.1.} 
For $s=2$, the lower order terms $w_1^iw_2^j$ with $i,j \le 2$ read as follows:
\begin{align}
Q_1 &= 1 + w_1 - w_1^2 + w_2 - 3 w_1 w_2 + 10 w_1^2 w_2 - 3 w_2^2 + 
 20 w_1 w_2^2 - 105 w_1^2 w_2^2+ \cdots,
\\
Q_2 &=1 + w_1 - w_1^2 + 2 w_2 - 4 w_1 w_2 + 12 w_1^2 w_2 - 
 5 w_2^2 + 28 w_1 w_2^2 - 135 w_1^2 w_2^2+ \cdots,
\\
w_1Y_1 &= 1 + w_1 - w_1^2 + 2 w_2 - 4 w_1 w_2 + 12 w_1^2 w_2 - 
 5 w_2^2 + 28 w_1 w_2^2 - 135 w_1^2 w_2^2+ \cdots,
\\
w_2Y_2 &= 1 + 2 w_1 - w_1^2 + 3 w_2 - 4 w_1 w_2 + 12 w_1^2 w_2 - 
 6 w_2^2 + 30 w_1 w_2^2 - 140 w_1^2 w_2^2+ \cdots,
\\
\rho_1 &= w_1-3 w_1^2-4 w_1 w_2+30 w_1^2 w_2+27 w_1 w_2^2-308 w_1^2 w_2^2
+ \cdots,
\\
\rho_2 &=w_2-4 w_1 w_2+15 w_1^2 w_2-7 w_2^2+54 w_1 w_2^2-308 w_1^2 w_2^2
+ \cdots,
\\
\varepsilon_1 &= 
w_1-3 w_1^2+w_2-8 w_1 w_2+45 w_1^2 w_2-7 w_2^2+81 w_1 w_2^2-616 w_1^2 w_2^2
+ \cdots,
\\
\varepsilon_2 &= w_1-3 w_1^2+2 w_2-12 w_1 w_2+60 w_1^2 w_2-14 w_2^2+135 w_1 w_2^2-924 w_1^2 w_2^2+ \cdots.
\end{align}
%\end{example}

The results (\ref{deema0}), (\ref{deema})  and (\ref{fq1}) indicate
\begin{align}\label{tkoshi}
\text{$Q_1$ for GGE$(\beta_1,\ldots, \beta_r, 0,\ldots,0,\beta_s)$}\;
\overset{s\rightarrow \infty}{\longrightarrow} \; 
\text{Largest eigenvalue of $V$ in (\ref{deema0})},
\end{align}
where the power series expansion of the LHS is obtained 
by specializing (\ref{qmu}) to $\mu_k = \delta_{k,1}$.

%\begin{example}\label{ex:b2}
\par \vspace{0.3cm}\noindent{\bf Example B.2.} 
The power series (\ref{qmu})  for $Q_1$ 
for GGE$(\beta_1,\beta_2, 0,\ldots,0,\beta_s)$ converges 
as $s$ gets large. 
In terms of the variables $z_j=\mathrm{e}^{-\beta_j}\,(j=1,2,\infty)$, 
the lower order part of the limit is
\begin{align}
Q_1&=1+z_1 z_2 z_\infty+z_1 z_2^2 z_\infty^2-z_1^2 z_2^2 z_\infty^2
+z_1 z_2^2 z_\infty^3-3 z_1^2 z_2^3 z_\infty^3+2 z_1^3 z_2^3 z_\infty^3
\\
&+z_1 z_2^2 z_\infty^4-3 z_1^2 z_2^3 z_\infty^4-3 z_1^2 z_2^4 z_\infty^4
+10 z_1^3 z_2^4 z_\infty^4-5 z_1^4 z_2^4 z_\infty^4+ \cdots
\end{align}
up to the terms $z_1^iz_2^jz_\infty^k$ with $\min(i,j,k) \ge 1$.
One can check that this indeed gives the 
largest eigenvalue of $V$ in (\ref{vmat})  confirming (\ref{tkoshi}).
For instance when $(z_1,z_2,z_\infty) = (0.4,0.3,0.2)$, 
they take the value 1.02511....
%\end{example}

%--------------------------------------------------
\section{\mathversion{bold}Proof of (\ref{kawa})}\label{app:note29}
%--------------------------------------------------

We illustrate the proof partly along the $4 \times 4$ example 
\begin{align}
{\footnotesize
\begin{pmatrix}
p_1+2m_1 & 2m_2  & 2m_3 & 2m_4 
\\
2m_1 & p_2+4m_2  & 4m_3 & 4m_4 
\\
2m_1 & 4m_2 & p_3+6m_3 & 6m_4
\\
2m_1 & 4m_2 & 6m_3 & p_4+ 8m_4
\end{pmatrix}
\begin{pmatrix}
x_1 & x_1 & x_1 & x_1
\\
x_1 & x_2 & x_2 & x_2
\\
x_1 & x_2 & x_3 & x_3
\\
x_1 & x_2 & x_3 & x_4
\end{pmatrix} = -2 
\begin{pmatrix}
p_1^{-1} & p_2^{-1} & p_3^{-1} & p_4^{-1}
\\
p_1^{-1} & 2p_2^{-1} & 2p_3^{-1} & 2p_4^{-1}
\\
p_1^{-1} & 2p_2^{-1} & 3p_3^{-1} & 3p_4^{-1}
\\
p_1^{-1} & 2p_2^{-1} & 3p_3^{-1} & 4p_4^{-1}
\end{pmatrix}.
}
\end{align}
From each row subtract the adjacent upper row starting from the bottom.
The result reads
\begin{align}\label{tats}
{\footnotesize
\begin{pmatrix}
p_1+2m_1 & 2m_2  & 2m_3 & 2m_4 
\\
-p_1 & p_2+2m_2  & 2m_3 & 2m_4 
\\
0 & -p_2 & p_3+2m_3 & 2m_4
\\
0 & 0 & -p_3 & p_4+ 2m_4
\end{pmatrix}
\begin{pmatrix}
x_1 & x_1 & x_1 & x_1
\\
x_1 & x_2 & x_2 & x_2
\\
x_1 & x_2 & x_3 & x_3
\\
x_1 & x_2 & x_3 & x_4
\end{pmatrix} = -2 
\begin{pmatrix}
p_1^{-1} & p_2^{-1} & p_3^{-1} & p_4^{-1}
\\
0 & p_2^{-1} & p_3^{-1} & p_4^{-1}
\\
0 & 0 & p_3^{-1} & p_4^{-1}
\\
0 & 0 & 0 & p_4^{-1}
\end{pmatrix}.
}
\end{align}
From $p_i = L-2\sum_{j=1}^4 \min(i,j)m_j$, one has the relations 
\begin{equation}\begin{split}
&p_1+2m_1+ 2m_2+2m_3+2m_4 = p_0,\qquad\;
p_2+2m_2+2m_3+2m_4 = p_1,\\
&p_3+2m_3+2m_4 = p_2,\qquad\qquad\qquad\qquad
p_4+2m_4=p_3,
\end{split}
\end{equation}
where $p_0=L$.
Thus taking the matrix product in (\ref{tats}) leads to the equation of the form
\begin{align}
\begin{pmatrix}
p_0 x_1 & s_{12}  & s_{13} & s_{14}
\\
0 & -p_1x_1+p_1x_2 & s_{23} & s_{24} 
\\
0 & 0  & -p_2x_2+p_2x_3 & s_{34}
\\
0 & 0 & 0 & -p_3x_3+ p_3x_4
\end{pmatrix}
= -2 
\begin{pmatrix}
p_1^{-1} & p_2^{-1} & p_3^{-1} & p_4^{-1}
\\
0 & p_2^{-1} & p_3^{-1} & p_4^{-1}
\\
0 & 0 & p_3^{-1} & p_4^{-1}
\\
0 & 0 & 0 & p_4^{-1}
\end{pmatrix}
\end{align}
for $s_{ij}$ which will be given explicitly later.
Thus the lower triangular part of the matrix equation is automatically satisfied.
The diagonal part compels (\ref{kawa}), i.e., 
\begin{align}\label{ymk}
x_k = -2\sum_{j=1}^k\frac{1}{p_{j-1}p_j}
\end{align}
as a {\em necessary} condition.
For sufficiency, one needs to further verify that (\ref{ymk}) 
also guarantees the equalities 
\begin{align}\label{sijp}
s_{ij} &= -\frac{2}{p_j}\quad (i<j).
\end{align}
Let us write down the element $s_{ij}$ for the general 
size $g\times g$ case.
By imagining the equation (\ref{tats}) for such a situation we have
\begin{align}\label{toksan}
s_{ij} = -p_{i-1}x_{i-1}+(p_i+2m_i)x_i + 2(m_{i+1}x_{i+1}+\cdots + m_{j-1}x_{j-1})
+2(m_j+\cdots +m_g)x_j,
\end{align}
where $1\le i< j \le g$ and $x_0:=0$.
From $p_i = L-2\sum_{j=1}^g \min(i,j)m_j$, one has  
\begin{align}
2m_k = p_{k-1}-2p_k+p_{k+1} \,(1\le k < g),\qquad 2m_g = p_{g-1}-p_g.
\end{align}
Substitution of them into (\ref{toksan}) gives
\begin{align}
s_{ij} &= -p_{i-1}x_{i-1} + (p_{i-1}-p_i+p_{i+1})x_i +
\sum_{i+1\le k \le j-1}(p_{k-1}-2p_k+p_{k+1})x_k + (p_{j-1}-p_j)x_j
\nonumber\\
&= p_{i-1}(x_i-x_{i-1}) +  p_{i}(x_{i+1}-x_{i})
+ \sum_{i+1\le k \le j-1}p_k(x_{k-1}-2x_k+x_{k+1}) + p_j(x_{j-1}-x_j).
\label{spx}
\end{align}
Now that $x_k$ dependence enters only through the difference 
\begin{align}
x_k-x_{k-1} = \frac{-2}{p_{k-1}p_k}
\end{align}
implied by (\ref{ymk}), 
the quantity (\ref{spx}) can be expressed entirely by $p_0,\ldots, p_g$.
After many cancellations, one finds that the result exactly yields $-\frac{2}{p_j}$,
completing a proof of (\ref{sijp}).

%--------------------------------------------------
\section{\mathversion{bold}Current in GGE$(\beta_1, \beta_\infty)$}\label{app:current1}
%--------------------------------------------------

The general result (\ref{Jl}) and (\ref{vsig}) 
on the current and the effective speed can be 
evaluated explicitly in the GGE$(\beta_1, \beta_\infty)$ 
treated in Sec. \ref{ss:2gge}.
In fact from (\ref{sigaz}) we have
\begin{align}
\frac{1}{\sigma_{j-1}\sigma_j} = 
\frac{(1+a)^2}{(1-a)^2}+
\frac{2(1+a)^2(1+z)}{(1-a)^2(1-z)}
\Bigl(\frac{1}{1+az^{j-1}}-\frac{1}{1+az^j}\Bigr).
\end{align}
Thus the sum (\ref{vsig}) can be taken, yielding the effective speed:
\begin{align}
v^{(l)}_k &= \frac{1+az^{l}}{1-az^{l}}v_{\min(k,l)},
\qquad 
v_k = \frac{1+a}{1-a}k - \frac{2a(1+z)(1-z^k)}{(1-a)(1-z)(1+az^{k})}.
\label{mirei}
\end{align}
Further substituting this and (\ref{roaz}) into (\ref{Jl}), 
we obtain, after some calculation, the stationary current:
\begin{align}\label{chizuko}
J^{(l)} &= 
\frac{a(1+z)}{(1+a)(1-z)}\Bigl(
1-\frac{(1-a)z^l}{1-az^l}\Bigr)-\frac{l a z^l}{1-az^l}.
\end{align}
From (\ref{qlim}),  one deduces some typical behavior as
\begin{alignat}{2}
J^{(l)} & = \frac{z(1-z^l-lz^l(1-z))}{(1-z)(1-z^{l+1})} 
+ O(\beta_1) \quad &(\beta_1\rightarrow 0),
\\
&= \frac{z((1+z)(1-z^l)-lz^l(1-z)}{(1-z)^3}\mathrm{e}^{-\beta_1}
+O(\mathrm{e}^{-2\beta_1})\quad &(\beta_1 \rightarrow \infty), 
\\
&= \frac{l}{2}- 
\frac{l(-1+3y^{-1}+3ly^{-1/2}+l^2)\beta_\infty}
{12(y^{-1/2}+l)}+O(\beta_\infty^3)
\quad & (\beta_\infty \rightarrow 0),
\\
&= \mathrm{e}^{-\beta_1-\beta_\infty} 
+ O(\mathrm{e}^{-2\beta_\infty})
\quad & (\beta_\infty \rightarrow \infty),
\end{alignat}
where $y$ and $z$ are defined in (\ref{ydef}).

\vspace{0.4cm}

%--------------------------------------------------
\section{\mathversion{bold}Alternative derivation of $J^{(l)}$
 in GGE$(\beta_1, \beta_\infty)$}\label{app:current2}
%--------------------------------------------------
Let us rederive the current (\ref{chizuko}) by an independent method
as a consistency check.
Consider the concatenation of two vertices from (\ref{vd}) 
which forms a segment in the 
diagram for the time evolution $T_l$ in  (\ref{tl}) as
\begin{equation}\label{figT}
\begin{picture}(100,35)(36,-11)
\multiput(0,0)(43,0){2}{
\put(14,10.5){\vector(0,-1){24}}\put(0,0){\vector(1,0){28}}}
\put(11,14.5){$\alpha$}\put(54,14.5){$\beta$}
\put(-9,-2){$n$}\put(32,-2){$m$}
\put(95,-2){$n,m \in \{0,1,\ldots, l\}, \quad\alpha, \beta \in \{0,1 \}.$}
\end{picture}
\end{equation}
Introduce the local transfer matrix $T$ generating the Boltzmann weight of this configuration 
in GGE$(\beta_1, \beta_\infty)$ as 
\begin{align}\label{Tdef}
T|n, \alpha\rangle = \sum_{m,\beta} y^{\theta(\alpha<\beta)}z^\beta |m,\beta\rangle,
\end{align}
where the parameters $y$ and $z$ are specified by
\begin{align}\label{ydef}
 y= \mathrm{e}^{-\beta_1} = \left(\frac{z^{\frac{1}{2}}-z^{-\frac{1}{2}}}
{a^{\frac{1}{2}}-a^{-\frac{1}{2}}}\right)^2,\qquad
z = \mathrm{e}^{-\beta_\infty}
\end{align}
according to (\ref{qlim}),  and the local forms
of the energies $E_1$ (\ref{e1}) and $E_\infty$ (\ref{ei}) have been taken into account.
Explicitly (\ref{Tdef}) reads as
\begin{align}
T|n,0\rangle &= |n-1,0\rangle + yz |n-1,1\rangle\quad (1 \le n \le l),
\\
T|0,0\rangle &= |0,0\rangle + yz |0,1\rangle,
\\
T|n,1\rangle &= |n+1,0\rangle + z|n+1,1\rangle\quad (0 \le n \le l-1),
\\
T|l,1\rangle &= |l,0\rangle + z |l,1\rangle.
\end{align}
The action on the dual basis is defined by postulating 
$(\langle n,\alpha|T)|n',\alpha'\rangle 
= \langle n, \alpha|  (T|n',\alpha'\rangle)$ 
and $\langle n,\alpha|n',\alpha'\rangle 
= \delta_{n,n'}\delta_{\alpha, \alpha'}$. 
Explicitly they read 
\begin{align}
\langle n,0|T &= \langle n+1, 0| + \langle n-1,1| \quad (1 \le n \le l-1),
\\
\langle 0,0|T &= \langle 0, 0| + \langle 1,0|,
\\
\langle l,0|T &= \langle l, 1| + \langle l-1,1|,
\\
\langle n,1|T &= yz \langle n+1, 0| + z \langle n-1,1| \quad (1 \le n \le l-1),
\\
\langle 0,1|T &= yz \langle 1, 0| + yz \langle 0,0|,
\\
\langle l,1|T &= z \langle l-1, 1| + z \langle l,1|.
\end{align}
We stay in the regime $0<a\le z <1$ as mentioned after (\ref{qlim}).
Then the left and right Perron-Frobenius eigenvectors 
has the eigenvalue $\frac{1-az}{1-a}$, and they are given by
\begin{align}
|\psi\rangle &= \frac{z(1-a)}{a(1-z)}|0,0\rangle+ 
\sum_{n=1}^l z^n |n,0\rangle + \frac{z(1-z)}{1-a}\sum_{n=0}^{l-1}z^n |n,1\rangle
+z^{l+1}|l, ,1\rangle,
\\
\langle \bar{\psi}| & = \sum_{n=0}^l\langle n,0| + \frac{1-a}{1-z}\sum_{n=0}^l \langle n,1|.
\end{align}
By writing them as $|\psi\rangle = \sum_{n,\alpha}\psi_{n,\alpha}|n,\alpha\rangle$ and 
$\langle \bar{\psi}| = \sum_{n,\alpha} \bar{\psi}_{n,\alpha} \langle n,\alpha|$,
the probability $\Bbb{P}(n,\alpha)$ of having the $(n,\alpha)$ part of the configuration in (\ref{figT}) is
\begin{align}
\Bbb{P}(n,\alpha) = \frac{\bar{\psi}_{n,\alpha}\psi_{n,\alpha}}
{\sum_{n,\alpha}\bar{\psi}_{n,\alpha}\psi_{n,\alpha}}.
\end{align}
A direct calculation of this gives
\begin{align}
\Bbb{P}(0,0) &= \frac{1-a}{(1+a)(1-a z^l)},
\quad
\Bbb{P}(n,0) = \frac{a(1-z)z^{n-1}}{(1+a)(1-a z^l)}\;\;(1 \le n \le l),
\\
\Bbb{P}(l,1)  &= \frac{a(1-a)z^l}{(1+a)(1-az^l)},\quad
\Bbb{P}(n,1) = \frac{a(1-z)z^{n}}{(1+a)(1-a z^l)}\;\;(0 \le n \le l-1).
\end{align}
The probability of the capacity $l$ carrier for holding $n$ balls is
\begin{align}
\Bbb{P}(n)  = \Bbb{P}(n,0) + \Bbb{P}(n,1) = \begin{cases}
\frac{1-az}{(1+a)(1-az^l)} & (n=0),
\\
\frac{az^{n-1}(1-z^2)}{(1+a)(1-az^l)} & (1 \le n \le l-1),
\\
\frac{az^{l-1}(1-az)}{(1+a)(1-az^l)} & (n=l).
\end{cases}
\end{align}
When $l \rightarrow \infty$ this result agrees with \cite[Lem.~3.15]{CKST18}
by identifying the parameter $p_0, p_1$ therein as 
$p_0 = \frac{a(1-z)}{1-az}, \,p_1 = \frac{z(1-a)}{1-a z}$.
Now it is elementary to calculate the expectation value of the number of balls in the carrier as
\begin{align}
\sum_{n=1}^l n \Bbb{P}(n) = 
\frac{a(1+z)}{(1+a)(1-z)}\Bigl(
1-\frac{(1-a)z^l}{1-az^l}\Bigr)-\frac{l a z^l}{1-az^l}.
\label{npn}
\end{align}
This reproduces the current (\ref{chizuko}).

%--------------------------------------------------
\section{\mathversion{bold} Linearly degenerate hydrodynamic type systems}\label{app:degenerate}
%--------------------------------------------------

A junction point between GHD and  linearly degenerate hydrodynamic type systems
\cite{Tsarev_1991, Kamchatnov_nonlinear_2000, Bulchandani2017}
can be constructed through the current conservation (\ref{cons}):
\begin{equation}
\partial_t \sigma+\partial_x(\sigma v)=0
\label{cons10}
\end{equation}
together with the characteristic equation (\ref{cons1}):
\begin{equation}
 \partial_t y +{\bf v}  \partial_x y=0.
\label{cons11}
\end{equation}
We now take into account that the densities $\sigma$ and the velocities $v$ are functions of the vector $y$, and this
vector $y$ is a function of $x$ and $t$. We can thus write $\partial_t \sigma_i=\sum_p \partial_t (y_p)\partial_p \sigma_i$, where $\partial_p$ means $\partial_p= \frac{\partial }{\partial y_p}$.
Similarly,  $\partial_x (\sigma_i v_i)=\sum_p \partial_x (y_p)\left[ v_i \partial_p \sigma_i + \sigma_i \partial_p v_i\right]$. Combined with $\partial_t (y_p) + v_p \partial_x (y_p)=0$, 
(\ref{cons10}) becomes:
\begin{equation}
\sum_p \left\{ - v_p \partial_x (y_p)\partial_p \sigma_i + \partial_x (y_p)\left[ v_i \partial_p \sigma_i + \sigma_i \partial_p v_i\right]\right\}=0.
\end{equation}
Since this relation should be verified for any vector $\partial_x y$ (one can choose arbitrarily the initial condition), we  have
\begin{equation}
\forall i,p\;\;\;v_p\partial_p \sigma_i  = v_i \partial_p \sigma_i + \sigma_i \partial_p v_i,
\end{equation}
or, equivalently:
\begin{equation}
{\partial_p \log(\sigma_i)}={\partial_p v_i\over v_p-v_i }.
\label{comp1}
\end{equation}
In particular, setting $p=i$, we find that $v_i$ does not depend on $y_i$:
\begin{equation}
{\partial_i v_i=0 }\;\;{\rm (no\;summation\;on\;} i{\rm)}.
\label{hydro0}
\end{equation}

Conversely, we can use the characteristic equation (\ref{cons11}) as a starting point and require the integrability conditions:
\begin{equation}
{\partial_k} ({\partial_p v_i\over v_p-v_i })={\partial_p} ({\partial_k v_i\over v_k-v_i }),\  p\ne k\ne i,
\label{hydro1}
\end{equation}
called  ``semi-Hamiltonian'' property, together with (\ref{hydro0}) called ``linear degeneracy''. The densities $\sigma_k$ defining the conserved
currents (\ref{cons10}) are then obtained by integrating (\ref{hydro1}).

In the BBS case, the $y$ dependence of the velocities follows from the equation (\ref{vitess2}):
\begin{equation}
v=\kappa^{dr}/1^{dr}.
\label{hydro10}
\end{equation}

We know from the commmutation of the transfer matrices $T_l$, $T_{l'}$ that the flows associated to two different sets of bare velocities $\kappa^{(l)}$ and $\kappa^{(l')}$ commute.
One can verify this propety directly within this formalism for an arbitrary pair $\kappa$, $\kappa'$ of vectors, by evaluating ${d\over dt}{dy_i\over dt'}-{d\over dt'}{dy_i\over dt}$.\footnote{ The time
variables $t$ and $t'$ are respectively associated to the dynamics with $\kappa$ and $\kappa'$.} A direct computation using (\ref{cons11}) shows it is equal to
\begin{equation}
 {d\over dt}{dy_i\over dt'}-{d\over dt'}{dy_i\over dt}
 =\sum_{p\ne i} \left(v'_p-v'_i\right)\left(v_p-v_i\right)
 \left[
 \frac{\partial_p v'_i}{v'_p-v'_i}-\frac{\partial_p v_i}{v_p-v_i}
 \right] \partial_x y_p\partial_x y_i.
\end{equation}
The term in brackets turns out to vanish due to (\ref{comp1}) and to the fact that $\sigma=1^{\rm dr}$ does not depend on the bare velocities ($\kappa$ or $\kappa'$). We thus get ${d\over dt}{dy_i\over dt'}={d\over dt'}{dy_i\over dt}$.

%\section*{Acknowledgments}
\ack
The authors thank Patrick Dorey, Dmytro Volin and Konstantin Zarembo for a kind interest.
Special thanks are due to 
Benjamin Doyon and Makiko Sasada for useful communications.
This work is partially supported by Osaka City University Advanced
Mathematical Institute (OCAMI)
and Grants-in-Aid for Scientific Research No.~16H03922, 18H01141 and
18K03452 from JSPS.

\end{document}